\documentstyle[11pt,amssymb]{article}

\makeatletter
\@addtoreset{equation}{section}
\makeatother

\makeatletter
\def\citen#1{%
\edef\@tempa{\@ignspaftercomma,#1, \@end, }
\edef\@tempa{\expandafter\@ignendcommas\@tempa\@end}%
\if@filesw \immediate \write \@auxout {\string \citation {\@tempa}}\fi
\@tempcntb\m@ne \let\@h@ld\relax \def\@citea{}%
\@for \@citeb:=\@tempa\do {\@cmpresscites}%
\@h@ld}
\def\@ignspaftercomma#1, {\ifx\@end#1\@empty\else
   #1,\expandafter\@ignspaftercomma\fi}
\def\@ignendcommas,#1,\@end{#1}
\def\@cmpresscites{%
 \expandafter\let \expandafter\@B@citeB \csname b@\@citeb \endcsname
 \ifx\@B@citeB\relax 
    \@h@ld\@citea\@tempcntb\m@ne{\bf ?}%
    \@warning {Citation `\@citeb ' on page \thepage \space undefined}%
 \else
    \@tempcnta\@tempcntb \advance\@tempcnta\@ne
    \setbox\z@\hbox\bgroup 
    \ifnum0<0\@B@citeB \relax
       \egroup \@tempcntb\@B@citeB \relax
       \else \egroup \@tempcntb\m@ne \fi
    \ifnum\@tempcnta=\@tempcntb 
       \ifx\@h@ld\relax 
          \edef \@h@ld{\@citea\@B@citeB }%
       \else 
          \edef\@h@ld{\hbox{--}\penalty\@highpenalty
            \@B@citeB }%
       \fi
    \else   
       \@h@ld\@citea\@B@citeB
       \let\@h@ld\relax
 \fi\fi%
 \def\@citea{,\penalty\@highpenalty\hskip.13em plus.1em minus.1em}%
}
\def\@citex[#1]#2{\@cite{\citen{#2}}{#1}}%
\def\@cite#1#2{\leavevmode\unskip
  \ifnum\lastpenalty=\z@\penalty\@highpenalty\fi
  \ [{\multiply\@highpenalty 3 #1
      \if@tempswa,\penalty\@highpenalty\ #2\fi 
    }]\spacefactor\@m}
\makeatother

\let\a=\alpha \let\b=\beta \let\g=\gamma \let\d=\delta \let\e=\epsilon
\let\z=\zeta \let\h=\eta \let\q=\theta  
  \let\n=\nu

\def\nn{\nonumber} \def\bd{\begin{document}} \def\ed{\end{document}}
\def\ds{\documentstyle} \let\fr=\frac \let\bl=\bigl \let\br=\bigr
\let\Br=\Bigr \let\Bl=\Bigl 
\let\bm=\bibitem
\let\na=\nabla
\let\pa=\partial \let\ov=\overline 
\newcommand{\be}{\begin{equation}} 
\newcommand{\ee}{\end{equation}} 
\def\ba{\begin{array}}
\def\ea{\end{array}}
\def\ft#1#2{{\textstyle{{\scriptstyle #1}\over {\scriptstyle #2}}}}
\def\fft#1#2{{#1 \over #2}}
\def\del{\partial}
\def\vp{\varphi}
\def\sst#1{{\scriptscriptstyle #1}}
\def\oneone{\rlap 1\mkern4mu{\rm l}}
\def\simequiv{\buildrel\sim\over=}
\def\td{\tilde}
\def\wtd{\widetilde}
\def\ie{{\it i.e.\ }}
\def\im{{\rm i}}
\def\dalemb#1#2{{\vbox{\hrule height .#2pt
        \hbox{\vrule width.#2pt height#1pt \kern#1pt
                \vrule width.#2pt}
        \hrule height.#2pt}}}
\def\square{\mathord{\dalemb{6.8}{7}\hbox{\hskip1pt}}}
\def\R{\rlap{\rm I}\mkern3mu{\rm R}}
\def\sR{\rlap{\hbox{$\scriptstyle\rm I$}}\mkern3mu{\hbox{$
\scriptstyle\rm R$}}}
\def\E{\rlap{\rm I}\mkern3mu{\rm E}}
\def\Z{\rlap{\sf Z}\mkern3mu{\sf Z}}
\def\0{{\sst{(0)}}}
\def\1{{\sst{(1)}}}
\def\2{{\sst{(2)}}}
\def\3{{\sst{(3)}}}
\def\4{{\sst{(4)}}}
\def\5{{\sst{(5)}}}
\def\6{{\sst{(6)}}}
\def\7{{\sst{(7)}}}
\def\8{{\sst{(8)}}}
\def\n{{\sst{(n)}}}
\def\v{{\cal V}}
\def\semi{\ltimes}
\def\hA{\hat{\cal A}}
\def\ns{{\sst {\rm NS}}}
\def\rr{{\sst {\rm RR}}}
\def\tH{{\widetilde H}}
\def\cA{{\cal A}}
\def\cF{{\cal F}}
\def\ep{{\epsilon}}
\def\mapright#1{\smash{\mathop{-\!\!\!-\!\!\!-\!\!\!-\!\!\!-\!\!\!
             \longrightarrow}\limits^{#1}}}
\def\maprightt#1#2{\smash{\mathop{-\!\!\!-\!\!\!-\!\!\!-\!\!\!-\!\!\!
             \longrightarrow}\limits^{#1}_{#2}}}
\def\tmap{{\mapright{\scriptstyle \rm T-dual}}}
\def\gmap#1#2{{\mapright{\scriptstyle SL(#1, \sR)_{#2}}}}
\def\cmap{{\maprightt{\scriptstyle \rm self-duality}{\scriptstyle
\rm  constraint}}}
\def\IIA{{\rm IIA}}
\def\IIB{{\rm IIB}}

\newcommand{\ho}[1]{$\, ^{#1}$}
\newcommand{\hoch}[1]{$\, ^{#1}$}
\newcommand{\bea}{\begin{eqnarray}} 
\newcommand{\eea}{\end{eqnarray}} 
\newcommand{\ra}{\rightarrow}
\newcommand{\lra}{\longrightarrow}
\newcommand{\Lra}{\Leftrightarrow}
\newcommand{\ap}{\alpha^\prime}
\newcommand{\bp}{\tilde \beta^\prime}
\newcommand{\tr}{{\rm tr} }
\newcommand{\Tr}{{\rm Tr} } 
\newcommand{\NP}{Nucl. Phys. }

\newcommand{\tamphys}{\it Center for Theoretical Physics,
Texas A\&M University, College Station, Texas 77843}
\newcommand{\ens}{\it Laboratoire de Physique Th\'eorique de l'\'Ecole
Normale Sup\'erieure\hoch{2,3}\\
24 Rue Lhomond - 75231 Paris CEDEX 05}

\newcommand{\auth}{I.V. Lavrinenko\hoch{\ddagger}, 
H. L\"u\hoch{\dagger}, C.N. Pope\hoch{\ddagger\star1}
and T.A. Tran\hoch{\ddagger}}

\begin{document}
\null
\vspace{-2cm}
\begin{flushright}
\hfill{CTP TAMU-26/98}\\
\hfill{LPTENS-98/25}\\
\hfill{SISSARef-70/98/EP}\\
\hfill{hep-th/9807006}\\
\hfill{June 1998}\\
\end{flushright}

\begin{center}
{\bf{\large U-duality as General Coordinate Transformations, and
Spacetime Geometry}}

\vspace{15pt}
\auth

\vspace{10pt}

{\hoch{\dagger}\ens}

\vspace{10pt}

{\hoch{\ddagger}\tamphys}

\vspace{10pt}

{\hoch{\star} \it SISSA, Via Beirut No. 2-4, 34013 Trieste, Italy\hoch{2}}

\vspace{15pt}

\underline{ABSTRACT}
\end{center}

     We show that the full global symmetry groups of all the
$D$-dimensional maximal supergravities can be described in terms of
the closure of the internal general coordinate transformations of the
toroidal compactifications of $D=11$ supergravity and of type IIB
supergravity, with type IIA/IIB T-duality providing an intertwining
between the two pictures.  At the quantum level, the part of the
U-duality group that corresponds to the surviving discretised internal
general coordinate transformations in a given picture leaves the
internal torus invariant, while the part that is not described by
internal general coordinate transformations can have the effect of
altering the size or shape of the internal torus.  For example,
M-theory compactified on a large torus $T^n$ can be related by duality
to a compactification on a small torus, if and only if $n\ge 3$.  We
also discuss related issues in the toroidal compactification of the
self-dual string to $D=4$.  An appendix includes the complete results
for the toroidal reduction of the bosonic sector of type IIB
supergravity to arbitrary dimensions $D\ge3$.

{\vfill\leftline{}\vfill
\vskip  10pt
\footnoterule
{\footnotesize  \hoch{1} Research supported in part by DOE 
Grant DE-FG03-95ER40917 \vskip  -12pt} \vskip 14pt
{\footnotesize \hoch{2} Supported in part by the
EC under TMR contract ERBFMRX-CT96-0045 \vskip -12pt} \vskip 14pt
{\footnotesize
        \hoch{3} Unit\'e Propre du Centre National de la Recherche
Scientifique, associ\'ee \`a l'\'Ecole Normale Sup\'erieure
\vskip -12pt} \vskip 10pt
{\footnotesize \hoch{\phantom{2}} et \`a l'Universit\'e de Paris-Sud
\vskip -12pt} \vskip 14pt}

\pagebreak
\setcounter{page}{1}

\section{Introduction}\label{intro}

   U-dualities are discrete subgroups of the global Cremmer-Julia (CJ)
symmetries of supergravities \cite{cjgroups}, which are conjectured to
be exact symmetries of string theory or M-theory \cite{ht,w}. The
origin of the CJ symmetries is in general rather deep, involving a
subtle interplay between the geometry of spacetime and the structure
of the gauge fields and their interactions.  The subtleties increase
as the dimension of the spacetime reduces.  For example, in the case
of the maximal supergravities obtained by toroidal dimensional
reduction from $D=11$, the $GL(1,\R)\sim O(1,1)$ and $GL(2,\R)$ CJ
symmetries in $D=10$ and $D=9$ can be simply understood as the
lower-dimensional interpretation of the general coordinate
transformations in the internal directions, corresponding to constant
linear transformations $\delta z^i = -\Lambda^i{}_j\, z^j$ of the
coordinates $z^i$ on the torus.\footnote{To be precise, the $\R$
factor of the $GL(11-D,\R)$ transformations parameterised by
$\Lambda^i{}_j$ does not, as it stands, leave the $D$-dimensional
Einstein-frame metric invariant, and one needs to include a
compensating transformation under the homogeneous rescaling
transformation of the Lagrangian in order to achieve this
\cite{cjlp1}.}  However, in $D\le 8$ the analogous residual coordinate
transformation symmetries in the toroidal reduction from $D=11$
account for only a part of the full CJ symmetry group, and its full
structure involves in addition transformations on the ``matter'' gauge
fields coming from the 3-form potential of $D=11$.   On the
other hand, from the viewpoint of a toroidal reduction of type IIB
supergravity there are always parts of the CJ groups that cannot be
understood as general coordinate transformations.

   It was observed in \cite{cllpst} that lower-dimensional CJ groups
could be used to perform transformations on $p$-brane solutions in
higher dimensions that alter their asymptotic or singularity
structures.  This was done by diagonally dimensionally reducing the
$p$-brane to an instanton, and acting on this with an $SL(2,\R)$
subgroup of the CJ group in this lower-dimensional Euclidean-signature
theory.  This transformation is capable of shifting the constant term
in the harmonic function describing the original $p$-brane solution,
and it is this shift that is responsible for altering the asymptotic
or singularity structure of the solution.  It is manifest that the
symmetry that is achieving such a change of the structure of the
solution must be more than merely a general coordinate transformation.
Indeed it is only by considering such instanton transformations in
$D\le8$ that the associated solutions can be mapped into ones with
altered singularity structures, and, as we observed previously, it is
in $D\le8$ that the CJ groups go beyond general coordinate
transformations.

    In this paper, we shall show that by combining the two
descriptions of the $D$-dimensional theory, corresponding to a
dimensional reduction either from $D=11$ or from the type IIB theory,
it is in fact possible to interpret any CJ symmetry transformation as
coming from a general coordinate transformation in the internal space
of the toroidal compactification in one or the other of the two
pictures.  Thus one can construct the entire CJ groups in all
dimensions as the closure of the internal coordinate transformations
of $D=11$ supergravity and type IIB supergravity, together with the
T-duality that relates the IIA and IIB theories.  In fact if a given
CJ symmetry transformation does not come from the internal general
coordinate transformations in one of the reduction pictures, then it
necessarily will in the other reduction picture.  An example of this
was already known in $D=9$, where it was shown that the $SL(2,\R)$
symmetry of type IIB, enlarged to $GL(2,\R)$ upon compactification to
$D=9$, could alternatively be understood, by virtue of IIA/IIB
T-duality, as the internal coordinate transformations of the 2-torus
compactification of $D=11$ supergravity \cite{schwarz}.  An
application of this approach is that any U-duality transformation on a
particular $p$-brane can be mapped, {\it via} a sequence of T-duality
transformations, into a general coordinate transformation in either
the type IIA or the type IIB theory.  Ideas along these lines were
indeed used in \cite{hyun,bps} for the purpose of effecting the same
changes to the harmonic functions in $p$-brane solutions that can also
be achieved by U-duality transformations \cite{cllpst}.

   If maximal supergravity in $D\le9$ is obtained by toroidal
reduction from $D=11$ then there will be a manifest $GL(11-D,\R)$
global symmetry coming from the internal linear transformations of the
$(11-D)$-torus.  On the other hand, if the same $D$-dimensional theory
is obtained by toroidal reduction from ten-dimensional type IIB
supergravity then there will instead be a manifest $GL(10-D,\R)$
global symmetry (at the level of the equations of motion in even
dimensions).  This $GL(10-D,\R)$ group is not, of course, simply a
subgroup of the previous $GL(11-D,\R)$ of the $D=11$ reduction.  In
fact, as we shall show, the closure of the two global symmetry groups
gives the full CJ symmetry group in dimension $D$.  A simple
non-trivial example is provided by the case of maximal supergravity in
$D=8$, for which the full CJ group is $SL(3,\R)\times SL(2,\R)$.  The
$SL(3,\R)$ factor is easily understood from the point of view of a
$T^3$ reduction from $D=11$, as a subgroup of the $GL(3,\R)$
general-coordinate symmetry.  From this viewpoint, however, the origin
of the $SL(2,\R)$ factor is more obscure; it is generated by an
interaction between gravity and the 3-form potential $A_\3$ in $D=11$,
giving rise to a dilatonic scalar and an axion in $D=8$.  This
dilatonic scalar parameterises the volume of the 3-torus, and,
together with the axion from $A_\3$, forms the $SL(2,\R)$-invariant
factor in the scalar Lagrangian.  If, on the other hand, we look at
the $D=8$ theory from the viewpoint of a $T^2$ reduction from the type
IIB theory, then the $SL(2,\R)$ factor in the CJ group is now easily
understood as a subgroup of the $GL(2,\R)$ symmetry from internal
coordinate transformations on $T^2$.  It is now instead the $SL(3,\R)$
factor whose origin is more obscure; it is an extension of the
$SL(2,\R)$ symmetry of the ten-dimensional type IIB theory.  By
combining the two viewpoints, together with the knowledge, coming from
T-duality, that the two eight-dimensional theories are really the
same, we can therefore understand both the $SL(3,\R)$ and the
$SL(2,\R)$ factors of the CJ symmetry as coming from certain internal
coordinate transformations.

     It should be emphasised that it is in $D\le8$ that the
interpretation of the CJ symmetries as internal general coordinate
transformations becomes non-trivial, since only then is there a {\it
necessity} of combining the pictures from the $D=11$ and the type IIB
compacifications in order to be able to give such an interpretation
for the entire CJ groups.  In dimensions lower than eight the story is
a little more complicated than the one described above in $D=8$, since
we now have CJ groups that are simple, rather than products of two
factors, and the way in which the two sets of internal coordinate
transformations close on the full group is more involved.  In the rest
of this paper we shall develop a general formalism for showing how the
general coordinate symmetries fit together in the two pictures.  In
section 2, we show by considering the Dynkin diagrams for the CJ
algebras, that we can indeed put the $GL(10-D,\R)$ symmetry from the
type IIB reduction route together with the $GL(11-D,\R)$ from the
$D=11$ reduction route to give the full $D$-dimensional CJ group.  We
also give, in section 3, a detailed decomposition of any CJ group
transformation in terms of a sequence of general-coordinate
transformations of $D=11$ and type IIB supergravities, intertwined by
type IIA/IIB T-duality transformations.

    Within a given picture, the CJ transformations that can be
described as internal general coordinate transformations are rather
trivial, in the sense that they are really just reparameterisation
symmetries.  We show in section 4 that the remainder of the CJ
transformations in the given picture have a more interesting
geometrical interpretation, in that they correspond to symmetries that
can change the size or the shape of the internal compactifying torus.
These kinds of transformation therefore arise in $D\le8$.  For
example, in these dimensions there is a duality symmetry relating
M-theory compactified on a given torus to M-theory compactifed on tori
of different sizes or shapes.  However this symmetry of M-theory
becomes less mysterious from the type IIB point of view, since it then
corresponds to internal general coordinate transformations of the
compactifying torus of the type IIB picture.

    In section 5, we give analogous discussions for the toroidal
reductions to $D=4$ of six-dimensional pure gravity and $N=1$
self-dual supergravity, and we obtain their global symmetry groups.
After presenting our conclusions in section 6, we include an appendix
in which the complete results for the toroidal dimensional reduction
of the bosonic sector of type IIB supergravity are derived, in all
dimensions $D\ge3$.  This extends previous results for the toroidal
reduction of $D=11$ supergravity \cite{lpsol,cjlp1}.  The type IIB
reduction has a new subtlety concerned with the handling of the
self-duality constraint on the 5-form field strength $H_\5$.  We show
that in the reduced $D$-dimensional theory the $GL(10-D,\R)$ symmetry
associated with the internal general coordinate transformations is a
symmetry only at the level of the equations of motion, when the
spacetime dimension is even, but it is a symmetry of the Lagrangian in
odd dimensions.  Our analysis shows that the entire $GL(10-D,\R)$
internal general-coordinate symmetry is preserved and is manifest.

\section{U-duality as general coordinate transformations}

     In appendix A, we derive the $D$-dimensional theory following
from the toroidal dimensional reduction of the bosonic sector of type
IIB supergravity.  We do this by starting from the enlarged
ten-dimensional theory introduced in \cite{bho}, where the five-form
field strength $H_\5$ is not constrained to be self-dual.  This allows
a Lagrangian formulation, given in (\ref{2blag}), whose equations of
motion reduce, after imposing $H_\5={*H_\5}$, to those of type IIB
supergravity.  In our dimensionally-reduced Lagrangians (\ref{dlag})
the $D$-dimensional consequences of this self-duality condition have
yet to be imposed; they are given, dimension by dimension, in
(\ref{lowersd}).  This formulation has the advantage of allowing the
full $GL(10-D,\R)$ global symmetry coming from the general coordinate
transformations on the compactifying $(10-D)$-torus to be manifest.
Unlike in $D=10$ itself, it is possible to write a spacetime-covariant
$D$-dimensional Lagrangian with the consequences of the
ten-dimensional self-duality condition already imposed. One way to do
this is by using (\ref{lowersd}) to eliminate higher-degree fields in
favour of their lower-degree duals.  Of course when $D$ is even this
cannot be done in a fully $GL(10-D,\R)$-covariant way, since half of
the fields of degree $D/2$ must be eliminated in favour of the other
half.  This can be done in a $GL(9-D,\R)$-covariant way \cite{cjlp1}.
Thus in odd dimensions, the Lagrangian for $D$-dimensional
supergravity has the full $GL(10-D,\R)$ symmetry manifest, while in
even dimensions the $GL(10-D,\R)$ is manifest only at the level of the
equations of motion.
     
    In $D\le9$, the maximal supergravities can be obtained from the
dimensional reduction of either $D=11$ supergravity or of
ten-dimensional type IIB supergravity, with the two formalisms being
related by field redefinitions that are the field-theoretic precursors
of type IIA/IIB string T-duality.  The full global symmetry of
$D$-dimensional maximal supergravity is the maximally non-compact form
of $E_{\sst{(11-D)}}$.  The scalar sectors of the Lagrangians describe
the cosets $E_{\sst{(11-D)}}/H_{\sst{(11-D)}}$, where $H_n$ is the
maximal compact subgroup of $E_n$.  In the dimensional reduction
scheme of \cite{lpsol,cjlp1}, which we are also using in this paper,
the scalar coset naturally arises in a Borel gauge, with the axionic
scalars (those that do not appear in exponentials) having associated
dilaton vectors that form the positive roots of $E_{\sst(11-D)}$
\cite{cjlp1,lpsweyl}.  This means that the $E_{\sst(11-D)}$ group can
easily be identified, by recognising the subset of dilaton vectors
associated with the simple roots.

    In the dimensional reduction route from $D=11$, the metric tensor
yields $D$-dimensional potentials ${\cal A}_\1^i$ and ${\cal A}^i_{\0
j}$, with associated dilaton vectors $\vec b_i'$ and $\vec b_{ij}'$;
while the 3-form potential yields gauge potentials $A_\3$, $A_{\2 i}$,
$A_{\1 ij}$ and $A_{\0ijk}$ with dilaton vectors $\vec a$, $\vec a_i$,
$\vec a_{ij}$ and $\vec a_{ijk}$ \cite{lpsol,cjlp1}.  Here, the
indices $i,j,\ldots$ label the directions on the compactifying
$(11-D)$-torus, running from $i=1$ to $i=11-D$.  (Note that we use
primes on the dilaton vectors $\vec{ b_i'}$ and $\vec b_{ij}'$ here,
to distinguish them from the dilaton vectors $\vec b_\a$ and $\vec
b_{\a\b}$ in the type IIB reduction in appendix \ref{app}.)  The
dilaton vectors ${\vec b_{ij}}'$ and $\vec a_{ijk}$ are the positive
roots of $E_{\sst(11-D)}$ in $D\ge 6$.  In $D=5,4$ and 3 they are
augmented by $-\vec a$, $-\vec a_i$, and $\{-\vec a_{ij}, -\vec
b_i'\}$ respectively, corresponding to the fact that in these
dimensions additional axions arise from the dualisation of
higher-degree fields $A_\3$, $A_{\2i}$ and $\{A_{\1ij}, \cA_\1^i \}$.
In all cases, one finds that the simple roots, in terms of which all
the others can be written as sums with non-negative integer
coefficients, are $\vec b_{i,i+1}'$ and $\vec a_{123}$, where $1\le i
\le 10-D$ \cite{cjlp1}. It is easily shown from their dot products
that they give the $E_{\sst(11-D)}$ Dynkin diagram:

\bigskip\bigskip

\centerline{
\begin{tabular}{ccccccccccccc}\\
 $\vec b_{12}'$& &$\vec b_{23}'$& &$\vec b_{34}'$& &$\vec b_{45}'$
& &$\vec b_{56}'$& &$\vec b_{67}'$& &$\vec b_{78}'$ \\
 o&---&o&---&o&---&o&---&o&---&o&---&o\\
 &   & &   &$|$&   & &   & &   & &   & \\
 &   & &   &o&   & &   & &   & &   & \\
 &   & &   &$\vec a_{123}$& & &   & &   & &   & \\
\end{tabular}}
\bigskip\bigskip

\centerline{Diag. 1: The $D=11$ dilaton vectors $\vec b_{i,i+1}'$
and $\vec a_{123}$ generate the $E_n$ Dynkin diagram.}
\bigskip\bigskip

   In using this diagram, it is to be understood that in a given
dimension $D$, only those simple roots whose indices are less than or
equal to $(11-D)$ are present.  Note that the simple roots $\vec
b_{i,i+1}'$ are those of the $SL(11-D,\R)$ subalgebra of the
$GL(11-D,\R)$ general coordinate transformations on the compactifying
$(11-D)$-torus.

   We can present the analogous discussion for the type IIB reductions
given in appendix \ref{app}.  In this case the axions are $\chi$,
${\cal A}^\a_{\0\b}$, $B_{\0\a\b\g\d}$, $A^\ns_{\0\a\b}$ and
$A^\rr_{\0\a\b}$ with dilaton vectors $\vec d$, $\vec b_{\a\b}$, $\vec
c_{\a\b\g\d}$, $\vec a^\ns_{\a\b}$ and $\vec a^\rr_{\a\b}$. In $D=4$
and $D=3$ these dilaton vectors will be augmented by $\{-\vec a^\ns,
-\vec a^\rr\}$ and $\{-\vec a^\ns_\a, -\vec a^\rr_\a\}$ respectively,
corresponding to the extra axions coming from the dualisation of the
associated higher-degree fields.  Again, we find that the dilaton
vectors form the positive roots of $E_{\sst (11-D)}$, and here the
simple roots turn out to be $\vec d$, $\vec a^\ns_{23}$ and $\vec
b_{\a,\a+1}$, for $1\le\a\le 9-D$.  From their dot products we find
that they generate the following Dynkin diagram:
\bigskip\bigskip

\centerline{
\begin{tabular}{ccccccccccccc}\\
 $\vec d$& &$\vec a^\ns_{23}$& &$\vec b_{34}$& &$\vec b_{45}$
& &$\vec b_{56}$& &$\vec b_{67}$& &$\vec b_{78}$ \\
 o&---&o&---&o&---&o&---&o&---&o&---&o\\
 &   & &   &$|$&   & &   & &   & &   & \\
 &   & &   &o&   & &   & &   & &   & \\
 &   & &   &$\vec b_{23}$& & &   & &   & &   & \\
\end{tabular}}
\bigskip\bigskip

\centerline{Diag. 2: The type IIB dilaton vectors $\vec d$, $\vec
a^\ns_{23}$ and $\vec b_{\a,\a+1}$ generate the $E_n$ Dynkin diagram.}
\bigskip\bigskip

   Here, in a given dimension $D$ only those simple roots with
indices less than or equal to $(10-D)$ are present.  Note that the
simple roots $\vec b_{\a,\a+1}$ are those of the $SL(10-D,\R)$
subalgebra of the $GL(10-D,\R)$ general coordinate transformations on
the compactifying $(10-D)$-torus.

    By comparing the two Dynkin diagrams, we see that each circle
representing a simple root is associated with a general-coordinate
transformation dilaton vector $\vec b_{i,i+1}'$ or $\vec b_{\a,\a+1}$
in at least one of the two diagrams.  This means that by putting
together the $GL(11-D,\R)$ from the $D=11$ picture, and the
$GL(10-D,\R)$ from the type IIB picture, one is able to generate the
entire Dynkin diagram for $E_{\sst{(11-D)}}$.  In other words, the
entire CJ symmetry of $D$-dimensional maximal supergravity can be
generated from general coordinate transformations on the compactifying
tori of the two pictures, which are related by T-duality.

     This way of patching together the $E_{\sst{(11-D)}}$ algebra
naturally explains the fact that $E_{\sst{(11-D)}}$ is a symmetry of
the Lagrangian in odd dimensions, but only of the equations of motion
in even dimensions.  This stems from the fact that the $GL(10-D,\R)$
general coordinate symmetry of the type IIB picture is itself a
symmetry only at the level of the equations of motion in even
dimensions (as we discussed earlier), owing to the self-duality
constraint on the ten-dimensional 5-form field strength.  Thus if $D$
is even, then if a given transformation within $E_{\sst{(11-D)}}$ is a
symmetry of the Lagrangian, then it must have its origin, in the
$D=11$ picture, as a general coordinate transformation of the
$(11-D)$-torus.  On the other hand, if it is a symmetry only of the
equations of motion, then in the type IIB picture it must originate
from a general coordinate transformation on the $(10-D)$-torus.  For
example, in $D=8$ the $SL(3,\R)$ factor in the CJ group, which is a
symmetry of the Lagrangian, is indeed a subgroup of the $GL(3,\R)$
coming from general coordinate transformations on the $T^3$
compactifying $D=11$.  On the other hand the $SL(2,\R)$ factor in the
CJ group, which corresponds to an electric/magnetic duality for
membranes, is a symmetry only of the equations of motion, and indeed
it originates from the general coordinate transformations of the $T^2$
compactifying the type IIB theory.

     The procedures described above can also be applied to the
$O(10-D,10-D)$ symmetry of the $N=1$ theory.  In this case,
corresponding to truncating out the R-R sectors of the type IIA and
IIB theories, we observe that the Dynkin diagrams given in Diag.\ 1
and Diag.\ 2 lose their left-hand-most circles, so that they are now
the Dynkin diagrams for $O(10-D,10-D)$.  The T-duality transformation
is now a symmetry of the theory, which interchanges the two ``ears''
of the Dynkin diagram.  Thus the $O(10-D,10-D)$ symmetry of the $N=1$
theory can also be viewed as the general coordinate symmetry
$GL(10-D,\R)$ intertwined with the discrete T-duality symmetry.

\section{Details of the decomposition}\label{details}

     In the previous section we have shown how the general coordinate
transformations on the internal compactifying torus in the $D=11$
picture and in the type IIB picture patch together to make the full CJ
group of $D$-dimensional maximal supergravity. Here, we give the
detailed decomposition of the global symmetry transformations as a
sequence of general coordinate transformations in the two tori,
intertwined with type IIA/IIB T-duality transformations.

     To do this, we first make a correspondence between the various
fields of the dimensionally-reduced $D=11$ and type IIB theories in
$D$ dimensions.  The simplest way to do this is to make the
correspondence first in $D=9$; this is summarised in Table 1 below.

\bigskip\bigskip
\begin{center}
\begin{tabular}{|c|c|c|c|c|c|}\hline
    &\multicolumn{2}{|c|}{IIA} &
    &\multicolumn{2}{c|}{IIB} \\ \cline{2-6}
    & $D=10$ & $D=9$ &T-duality & $D=9$ & $D=10$ \\ \hline\hline
    & $A_\3$ & $A_\3$ & $\longleftrightarrow$ &
                   $B_{\3 2}$ & $B_\4$ \\ \cline{3-6}
R-R & &  ${A_{\2 2}}$& $\longleftrightarrow$
                           & $A_\2^{\rr}$ & $A_\2^{\rr}$
                                               \\ \cline{2-5}
fields& ${\cal A}_\1^{1}$ & ${\cal A}_\1^{1}$ &
                $\longleftrightarrow$ &
        ${A_{\1 2}^{\rr}}$ & \\ \cline{3-6}
   & & ${{\cal A}^1_{\0 2}}$ & $\longleftrightarrow$
                            & $\chi$ &$\chi$
                                 \\ \hline\hline
NS-NS & $G_{\mu\nu}$ & ${{\cal A}_\1^{2}}$
                        & $\longleftrightarrow$ &
        ${A_{\1 2}^{\ns}}$ & $A_\2^{\ns}$ \\ \cline{2-5}
fields& $A_{\2 1}$ & $A_{\2 1}$ &
               $\longleftrightarrow$ & $A_\2^{\ns}$ &
                                      \\ \cline{3-6}
      & & ${A_{\1 12}}$ & $\longleftrightarrow$ &
                              ${{\cal A}_\1^2}$ & $G_{\mu\nu}$
                                       \\ \hline
\end{tabular}
\end{center}

\bigskip

\centerline{Table 1: Gauge potentials of type II theories in $D=10$
and $D=9$}
\bigskip\bigskip

    From these correspondences in $D=9$, those in all $D\le 8$ can be
obtained by simply performing standard dimensional reductions from
$D=9$.  We now therefore introduce indices $a,b,\ldots$, running from
3 to $11-D$ (so that $i=(1,\a)=(1,2,a)$, {\it etc}.).  In $D$
dimensions we therefore have for the NS-NS fields

\bigskip

\centerline{
\begin{tabular}{cccccc}
IIA: & $A_{\2 1}$ & $A_{\1 12}$ & $A_{\1 1 a}$ & $A_{\0 12a}$ &
                                             $ A_{\0 1ab}$ \\
IIB: & $A_{\2}^\ns$ & $\cA_\1^2$ &$A^\ns_{\1a}$&$\cA^2_{\0a}$ &
                                             $A^\ns_{\0ab}$ \\
            \\
IIA: & $\cA^2_\1$ & $ \cA^a_\1$& $\cA^2_{\0a}$ & $\cA^a_{\0b}$ & \\
IIB: &$A^\ns_{\1 2}$ & $ \cA^a_\1$ & $A^\ns_{\02a}$&$\cA^a_{\0b}$ &
\end{tabular}}
\bigskip

{\noindent} and for the R-R fields

\bigskip
\centerline{
\begin{tabular}{cccccc}
IIA: & $A_{\3}$ & $A_{\2 2}$ & $A_{\2 a}$ & $A_{\1 2a}$ &
                                             $ A_{\1 ab}$ \\
IIB: & $B_{\3 2}^\ns$ & $A^\rr_\2$ &$B_{\22a}$&$A^\rr_{\1a}$ &
                                             $B_{\12ab}$ \\
            \\
IIA: & $A_{\02ab}$ & $A_{\0abc}$& $\cA^1_\1$ & $\cA^{1}_{\02}$ &
                                            $\cA^1_{\0a}$ \\
IIB: &$A^\rr_{\0ab}$&$B_{\02abc}$ & $A^\rr_{\12}$&$ \chi$ &
                                            $A^\rr_{\02a}$
\end{tabular}}
\bigskip

    Since the internal directions labelled by 1 and by 2 have both
been singled out in this correspondence, we see only $GL(9-D,\R)$,
corresponding to the $a,b,\ldots$ indices, as a manifest symmetry.
Furthermore, we have made a specific choice in keeping the particular
components of the $B$ potentials indicated above. The remaining ones,
namely $B_{\3a}$, $B_{\2ab}$, $B_{\1abc}$ and $B_{\0abcd}$, have field
strengths that are related by Hodge dualisation to the field strengths
of those that we are keeping.  This way of truncating the
supernumerary fields was discussed in \cite{cjlp1}.  As we discussed
earlier, the theory actually has the full $GL(11-D,\R)$ general
coordinate symmetry in the $D=11$ compactification, and the full
$GL(10-D,\R)$ in the type IIB compactification.

    We are now in a position to show how the full set of CJ
transformations can be expressed purely in terms of general coordinate
transformations on the compactifying torus of one or other of the two
pictures.  Without loss of generality, we shall choose to express the
CJ transformations of the $D=11$ picture in terms of these general
coordinate transformations.  As usual, we can concentrate on the
scalar manifolds, since these realise the entire $E_{\sst{(11-D)}}$ CJ
groups.  The transformations associated with the Kaluza-Klein axions
${\cal A}^i_{\0 j}$ originate from the $GL(11-D,\R)$ of the general
coordinate transformations of the $(11-D)$-torus compactifying $D=11$
supergravity, and so these are already of the required
general-coordinate form.  The transformations associated with the
remaining axions are not of the general-coordinate type, as seen from
the $D=11$ reduction viewpoint.  However, we shall now show that they
{\it can} be expressed as general coordinate transformations when seen
from the type IIB reduction viewpoint.  For $D\ge 6$, the extra axions
are just the $A_{\0ijk}$ potentials.  They can be mapped into
Kaluza-Klein axions in the type IIB picture as follows:
\be
\matrix{\IIA & & \IIA & & \IIB\cr
A_{\0ijk} & \gmap{11-D}{\rm M} & A_{\012a} & \tmap & \cA^2_{\0a}\cr}
\ee

    In $D\le 5$, in addition to the axions described above, which are
handled analogously, there are further ones that come from the
dualisations of higher-degree fields.  These can be discussed
dimension by dimension.  In $D=5$, the extra axion is obtained from
dualising the 3-form potential $A_\3$.  Since the process of
dualisation is non-local on the potentials, we shall indicate the
sequence of correspondences on the field strengths here.
\be
\matrix{
\IIA & & \IIB && \IIB && \IIB \cr
{*F_\4}&\tmap & {*H_{\42}} &\cmap & H_{\1abcd} &\gmap{5}{\IIB} &
H_{\12abc}\cr
&&&&&&\cr
& & \IIA & & \IIA && \IIB \cr
&\tmap & F_{\1abc} & \gmap{6}{\rm M} & F_{\1 12a} &
\tmap &\cF^2_{\1a}\cr}
\ee
The step indicated by ``self-duality constraint'' makes use of the
lower-dimensional self-duality relations given in (\ref{lowersd}).

     In $D=4$, there are seven extra axions coming from the
dualisation of $A_{\2i}$.  The sequence of their correspondences to
the Kaluza-Klein axions of the type IIB reduction is given by
\be
\matrix{
\IIA & & \IIA && \IIB && \IIB \cr
{*F_{\3i}}&\gmap{7}{\rm M} & {*F_{\3a}} &\tmap & {*H_{\32a}}
&\gmap{6}{\IIB} & {*H_{\3ab}}\cr
&&&&&&\cr
& & \IIB & & \IIA && \IIA \cr
&\cmap & H_{\12abc} & \tmap & F_{\1 abc} &
\gmap{7}{\rm M} &F_{\112a}\cr
&&&&&&\cr
& & \IIB & &&&\cr
&\tmap & \cF^2_{\1a} & &  & &\cr}
\ee

In $D=3$, there are twenty-eight additional axions coming from the
dualisation of $A_{\1ij}$ and eight from the dualisation of
$\cA^i_\1$.  We consider the twenty-eight axions first, for which we
find that the sequence of their correspondence is given by
\be
\matrix{
\IIA & & \IIA && \IIB && \IIB \cr
{*F_{\2ij}}&\gmap{8}{\rm M} & {*F_{\2ab}} &\tmap & {*H_{\22ab}}
&\gmap{7}{\IIB} & {*H_{\2abc}}\cr
&&&&&&\cr
& & \IIB & & \IIA && \IIA \cr
&\cmap & H_{\12abc} & \tmap & F_{\1 abc} &
\gmap{8}{\rm M} &F_{\112a}\cr
&&&&&&\cr
& & \IIB & &&&\cr
&\tmap & \cF^2_{\1a} & &  & &\cr}\label{d3seq1}
\ee

     The sequence for the other eight axions is given by
\be
\matrix{
\IIA & & \IIA && \IIB && \IIB \cr
{*\cF^i_{\2}}&\gmap{8}{\rm M} & {*\cF^1_{\2}} &\tmap & {*F^\rr_{\22}}
&\gmap{7}{\IIB} & {*F^\rr_{\2a}}\cr
&&&&&&\cr
& & \IIA & & \IIB &&\cr
&\tmap & {*F_{\22a}} & \gmap{8}{\rm M} & {*F_{\2 ab}} &
\cdots\cdots&\cr}
\ee
where the dots indicate that the sequence then follows (\ref{d3seq1}),
joining it at the second stage of the first line.

   To summarise, in this section we have shown that all the axions
that do not directly arise as Kaluza-Klein axions from $D=11$
supergravity can be mapped, by a sequence of internal
general-coordinate symmetry transformations and type IIA/IIB T-duality
transformations, into Kaluza-Klein axions from type IIB supergravity.
A completely analogous converse discussion would allow one to
interpret all the axions of the type IIB reduction as Kaluza-Klein
axions.

\section{M-theory interpretation of U-duality}

     So far in the paper, we have been focussing on how the full CJ
groups may be interpreted in terms of internal general coordinate
transformations, either directly on the compactifying torus in the
picture ($D=11$ supergravity or type IIB supergravity) under
consideration, or else indirectly, in the sense that we first make a
T-duality transformation to the other picture, and interpret the CJ
transformation as an internal general coordinate transformation there.

    It is of interest also, however, to take a different approach, and
to study how the full set of CJ transformations act when we stay
entirely within one picture or the other.  Some of the CJ
transformations will simply be internal general coordinate
transformations, but the others, which are more interesting from this
point of view, do not have this geometrical interpretation.

    To study this, let us consider in more detail the $D=8$ example
that we discussed previously.  We shall consider it as arising from
the compactification of M-theory on a 3-torus.  As we discussed
previously, the $SL(3,\R)$ factor in the $SL(3,\R)\times SL(2,\R)$
Cremmer-Julia group comes from internal general coordinate
transformations on $T^3$, while the $SL(2,\R)$ factor is
electric/magnetic symmetry of the membrane \cite{izq}.  The scalar and
gauge fields in the theory, in the notation of \cite{lpsol,cjlp1}, are
as follows:
\bea
\hbox{Scalars}: && 
\vec\phi\ ,\qquad \cA^i_{\0 j}\ ,\qquad \chi\equiv A_{\0 123}\ ,\nn\\
\hbox{Vectors}: && \cA_\1^i \ ,\qquad A_{\1 ij}\ ,\label{list}\\
\hbox{Higher-rank}: && A_{\2 i} \ , \qquad A_\3\ ,\nn
\eea
where $1\le i\le 3$.  The Lagrangian is given by
\bea
{\cal L} &=& R{*\oneone} -\ft12 {*d\vec\phi }\wedge d\vec\phi -\ft12
e^{\vec a\cdot
\vec\phi}\, {*F_\4}\wedge F_\4 -\ft{1}{2} \sum_i
e^{\vec a_i\cdot \vec\phi}\, {*F_{\3i}}\wedge F_{\3 i}\nn\\
&&-\ft12 \sum_{i<j} e^{\vec a_{ij}\cdot \vec\phi}\, {*F_{\2ij}}
\wedge F_{\2 ij}
-\ft12 \sum_i e^{\vec b_i\cdot \vec\phi}\, {*{\cal F}_\2^i}\wedge
{\cal F}_\2^i
-\ft12  e^{\vec a_{123} \cdot\vec \phi}\,
{*d\chi}\wedge d\chi \nn\\
&&-\ft12 \sum_{i<j} e^{\vec b_{ij}\cdot \vec\phi}\,
{*{\cal F}_\1^i{}_j} \wedge {\cal F}^i_{\1 j} + \ft12 \chi\, 
dA_\3\wedge dA_\3\nn\\
&& -\Big( \ft16 dA_{\2i}\wedge dA_{\2j}\wedge A_{\1 k} +\ft12
dA_\3 \wedge dA_{\2 i}\wedge A_{\1 jk} \Big)\, \ep^{ijk}\ ,
\label{dgenlag}
\eea
where the ``dilaton vectors'' $\vec a$, $\vec a_i$, $\vec a_{ij}$,
$\vec a_{ijk}$, $\vec b_i$, $\vec b_{ij}$ are constants that
characterise the couplings of the dilatonic scalars $\vec \phi$ to the
various gauge fields; they are given in \cite{lpsol,cjlp1}.

      The $i,j,\ldots$ indices are in the fundamental representation
of $SL(3,\R)$, and so in particular the Kaluza-Klein vectors
$\cA_\1^i$ form a triplet of $SL(3,\R)$.  Similarly, the ``membrane
wrapping mode'' vectors $A_{\1 ij}$ form another triplet of
$SL(3,\R)$.  The $SL(3,\R)/O(3)$ part of the scalar manifold is
parameterised by the Kaluza-Klein axions $\cA^i_{\0 j}$, together with
the two linear combinations of the three dilatons $\vec \phi$ that do
not couple to $A_{\0 123}$.  It is convenient to express these two
linear combinations in terms of the three linearly-dependent vectors
$\vec\b_i = \vec b_i +\ft12 \vec a_{123}$, which have the property
that $\vec \b_i\cdot \vec a_{123}=0$.  In terms of these, we may write
the dilaton vectors for the Kaluza-Klein and winding-mode vectors as
\bea
\vec b_i &=& \vec \b_i -\ft12\vec a_{123}\ ,\nn\\
\vec a_{ij} &=& \vec \b_k +\ft12 \vec a_{123}\, \qquad
i\ne j\ne k \ne i\ .\label{abrels}
\eea
In fact $A_{\0 123}$, which we are calling $\chi$ for convenience,
together with the remaining orthogonal linear combination
$\varphi\equiv \ft12\vec a_{123}\cdot\vec\phi$ of the three dilatons
$\vec\phi$, paremeterises the $SL(2,\R)/O(2)$ part of the scalar
manifold.  Under $SL(2,\R)$ the Kaluza-Klein vectors $\{\cA_\1^1,
\cA_\1^2, \cA_\1^3\}$ pair off with the membrane-wrapping vectors
$\{A_{\1 23}, A_{\1 13}, A_{\1 12}\}$ respectively, to form three
doublets.  The field strength $F_\4$ and its dual form another doublet
under $SL(2,\R)$, and the remaining fields are invariant.

     The ansatz for the reduction of the metric, following the
notation of \cite{cjlp1}, is
\be
ds_{11}^2 = e^{\fft13 \vec g\cdot\vec\phi}\, ds_8^2 + \sum_{i=1}^3
 e^{2\vec\gamma_i\cdot\vec\phi}\, (dz^i + \cA_\1^i + \cA^i_{\0 j}\,
dz^j)^2\ ,
\ee
where the vectors $\vec g$ and $\vec f_i$ can be found in
\cite{cjlp1}.  The volume form of the physical 3-torus is therefore
given by $e^{(\vec\g_1 +\vec\g_2+\vec \g_3)\cdot\vec\phi}\, dz^1\wedge
dz^2\wedge dz^3$.  The exponential factor turns out to be $e^{-\fft12
\vec a_{123}\cdot\vec\phi}$, and hence the volume form is
$e^{-\varphi}\, dz^1\wedge dz^2\wedge dz^3$.  Since $\varphi$ is the
dilaton of the $SL(2,\R)/O(2)$ scalar manifold, and since $dz^1\wedge
dz^2\wedge dz^3$ is invariant under the $SL(3,\R)$ transformations
$\delta z^i= \Lambda^i{}_j\, z^j$, it follows that the volume of the
3-torus, namely
\be
V_3 = e^{-\varphi}\ ,
\ee
is $SL(3,\R)$ invariant, and furthermore is given in terms of the
dilaton $\varphi$ of the $SL(2,\R)$ factor in the $D=8$ CJ group.

     In the Einstein frame, the mass per unit $p$-volume for an
electric $p$-brane supported by a field strength with kinetic term of
the form $e^{\vec c\cdot\vec\phi}\, F^2$ is $M\sim e^{-\fft12 \vec
c\cdot\vec\phi}$ \cite{dkl,lptdual}.  From (\ref{dgenlag}) and
(\ref{abrels}), we therefore see that the masses of the particles
supported by the Kaluza-Klein vectors and the winding-mode vectors
depend on the $SL(2,\R)$ dilaton as follows:
\be
M_{\rm KK} \sim e^{\fft12\varphi}\ ,\qquad
M_{\rm WM} \sim e^{-\fft12\varphi}\ .
\ee
(We are assuming here, for simplicity, that the axion $\chi$ is zero.
We shall discuss the more general case below.)  Thus, as expected, we
see that the Kaluza-Klein particles become massless in the limit when
the volume of the 3-torus goes to infinity.\footnote{To be precise, in
order to see that the BPS black hole solitons supported by the
Kaluza-Klein vectors indeed describe the Kaluza-Klein particle states
coming from $D=11$, we should calculate their masses in the $D=11$
M-metric and not the $D=8$ Einstein metric.  In general, if two
metrics are related by $g^{\sst{(A)}}_{\mu\nu} = \Omega^2\,
g^{\sst{(B)}}_{\mu\nu}$, where $\Omega$ is a constant, then the masses
per unit $p$-volume of a $p$-brane are related by $m^{\sst{(A)}} =
\Omega^{-p-1}\, m^{\sst{(B)}}$ \cite{lptdual}.  Thus the mass of the
Kaluza-Klein particles in the M-metric is of the form
$e^{\fft12\varphi}\, e^{-\fft16\vec g\cdot\vec\phi} \sim
e^{\ft13\varphi}=V^{-1/3}$.  This is the expected scale dependence for
the Kaluza-Klein masses arising from compactification on a 3-torus.
Note that the winding-mode masses in the M-metric are of the form
$\sim V^{2/3}$, which is what one would expect for a membrane.
(Actually these analyses, like all analyses of this kind, are
guaranteed to ``work,'' since on closer inspection they turn out to be
nothing but the verification that the lower-dimensional theory is
indeed the dimensional reduction of the higher-dimensional one.  Thus
the success of such ``miracles'' is somewhat tainted by the fact that
they would equally well appear to operate even in cases where there is
no supersymmetry and no underlying fundamental theory!)}  (This is the
same phenomenon as the one discussed by Witten in \cite{w}, where the
Kaluza-Klein particles in the type IIA string become massless as the
dilaton grows large, leading to the opening out of the eleventh
dimension.)  On the other hand, if the volume of the 3-torus goes to
zero then the Kaluza-Klein particles become heavy, while the
winding-mode particles now become massless.  Thus the particle
spectrum of M-theory compactified on a small $T^3$ is identical to the
particle spectrum of M-theory compactified on a large $T^3$ of
reciprocal volume.  This implies that M-theory compactified on a large
$T^3$ is dual to M-theory compactified on a small $T^3$ of reciprocal
volume.  (Recall that we are taking $\chi=0$ here for simplicity.)

   It is not enough just to study the particle spectrum in order to
justify this duality conjecture.  Further supporting evidence can be
found from the spectra of the other $p$-branes in the theory.  For
example, the masses of the fundamental string and two D-strings,
measured in the Einstein metric, are dependent only on the shape of
the $T^3$, but not on the size.  Actually, the conjecture about the
equivalence of the compactification on a large $T^3$ and a small $T^3$
is really just equivalent to the conjecture of U-duality, and the
supporting evidence for the one conjecture is equivalent to that for
the other.  This is a consequence of the fact that the conjectured
$T^3$ duality symmetry is implied by the $Z_2$ Weyl group symmetry of
$SL(2,\Z)$ factor of the U-duality group in $D=8$.

    In the CJ group $SL(3,\R)\times SL(2,\R)$, the $SL(3,\R)$ subgroup
comes from internal general coordinate transformations. Upon
quantisation, it is discretised to $SL(3,\Z)$.  This discrete subgroup
has the feature that it leaves the size and shape of the compactifying
3-torus unchanged.  For this reason, different moduli which are
related by this $SL(3,\Z)$ symmetry transformation should be
identified.  On the other hand, the $SL(2,\Z)$ factor of the U-duality
group cannot all be regarded as general coordinate transformations of
the internal space from the 11-dimensional point of view.  (In fact an
$\R$ subgroup of $SL(2,\R)$ does originate from the internal general
coordinate transformations, as we shall discuss below.)  It is
generated by the axion $A_{\0123}$ (which does not parameterise any
modulus of the 3-torus) and the dilatonic scalar $\varphi$ that
measures the volume of the compactifying 3-torus.  This implies that
under this $SL(2,Z)$ transformation, the volume of the 3-torus ranges
from infinity to zero.  The $D=8$ U-duality symmetry implies that
M-theory compactified on $M_8\times T^3$ for various sizes of the
volume that are related each other by $SL(2,\Z)$ transformations are
dual to each other.  To be precise, if we start with M-theory
compactified on a 3-torus of volume $V_0$, and a value $\chi_0$ for
the axion, then it is dual to M-theory compactified on a 3-torus of
the same shape but with volume $V$, and an axion $\chi$, given by
\be
V= \fft1{c^2 \, V_0 + (c\, \chi_0 + d)^2\, V_0^{-1}}\ ,\qquad
\chi= \fft{(a\, \chi_0 + b)(c\, \chi_0 + d) + a\, c\, V_0^2}{ c^2\,
V_0^2 + (c\, \chi_0 +d)^2}\ ,\label{abcd}
\ee
where $a$, $b$, $c$ and $d$ are integers satisfying $a\, d-b\, c =1$.

   Classically, the $GL(3,\R) \sim \R\times SL(3,\R)$ of internal
general coordinate transformations (which also utilises a compensating
``trombone'' \cite{trombone} rescaling symmetry so as to leave the
lower-dimensional metric invariant \cite{cjlp1}) can change the volume
as well as the shape of the 3-torus.\footnote{ The $GL(11-D, \R)$
symmetry of the $D$-dimensional supergravity, which acts on the lower
dimensional fields, can be derived from the covariance of the
eleven-dimensional theory under constant linear coordinate
transformations in the internal $T^{\sst{11-D}}$.  However, if the
$GL(11-D, \R)$ symmetry is viewed just as a transformation on the
$D$-dimensional fields, it will now imply an active transformation on
the internal space, and hence will have the effect of changing the
size and shape of the internal torus, since in the dimensional
reduction procedure a specific configuration for the internal
coordinates is chosen.  Nevertheless, we shall still refer to this
$D$-dimensional $GL(11-D, \R)$ global symmetry as internal general
coordinate transformations.}  In particular, it is the $\R$ factor
that rescales the volume, while the $SL(3,\R)$ factor changes the
shape.  However, the scalar corresponding to this rescaling symmetry
has been ``reassigned'' to the $SL(2,\R)$ factor in the full
$SL(3,\R)\times SL(2,\R)$ CJ group, and so upon quantisation, where it
is conjectured that the U-duality group $SL(3,\Z)\times SL(2,\Z)$
survives, the volume-changing transformations (\ref{abcd}) survive
within the $SL(2,\Z)$ factor, and not the $SL(3,\Z)$ factor.  This
should be contrasted with the type IIA $D=10$ and $D=9$ cases, where
the volume-changing factors in the classical $GL(1,\R)$ and $GL(2,\R)$
CJ groups are completely broken at the quantum level.  This explains
why in these two cases the physics is changed by changing the radius
of the $S^1$ or the volume of the $T^2$.  (In the $D=10$ case the
weakly-coupled type IIA string is mapped to a $D=11$ theory as the
circle gets large \cite{w}, whilst in the $D=9$ case the type IIB
theory emerges as the volume of the $T^2$ shrinks to zero
\cite{schwarz}.)  In $D\le8$, on the other hand, the volume of the
compactifying torus {\it can} be changed by surviving U-duality
symmetries, and so in such cases one has duality symmetries relating
M-theory compactified on tori of different radii.

      In the above discussion, we said that M-theory on large a
3-torus is dual to M-theory on a small 3-torus with reciprocal volume,
provided that the Kaluza-Klein modes and membrane wrapping modes are
interchanged.  At first sight, this might seem to be in contradiction
with a fundamental membrane interpretation, where the Kaluza-Klein
modes are those of the membrane itself.  If we consider a fundamental
membrane in $D=11$, then its particle modes compactified on $T^3$
might be expected to have masses of the form
\be
m^2\sim x\, V^{-\ft23} + y\, V^{\ft43}\ ,\label{naive}
\ee
where $x$ and $y$ are related to the mode numbers for the Kaluza-Klein
and winding modes respectively, and they would also depend on the
parameters specifying the fixed shape of the 3-torus.  This mass
formula is obviously not invariant under the $V\rightarrow 1/V$
transformation.  However the analogy between strings and membranes is
not so close as to make the mass formula (\ref{naive}) justifiable,
since there is no sensible perturbative spectrum analysis for
membranes that are not wrapped around the torus.  Thus the formula
(\ref{naive}) really only arises in an approximate analysis where the
Kaluza-Klein contributions are treated as a small perturbation to a
large winding-mode contribution \cite{dipss}.  It would therefore be
inappropriate to expect such an approximate formula to be invariant
under the interchange of Kaluza-Klein and winding modes.

      So far we have considered the U-duality in $D=8$.  The
generalisation to lower dimensional U-duality is straightforward.  The
$SL(11-D,\Z)$ leaves the size and shape of the $(11-D)$-torus
invariant, whilst the parts of the U-duality transformations that are
not internal general coordinate transformations can change the size or
shape of the torus.  This implies that there is a duality symmetry of
M-theory compactified on various sizes and shapes of internal torus.
To summarise, when M-theory is compactified on $M_{10}\times S^1$, the
small circle limit gives rise to type IIA perturbative string theory.
If $M_{11}=M_{9}\times T^2$, the small volume limit of the $T^2$ give
rise to the type IIB string.  If $M_{11}=M_8\times T^3$, there is an
$SL(2,\Z)$ multiplet of sizes for the volume that should be identified
(as given in (\ref{abcd})), but different shapes of the $T^3$ give
rise to inequivalent moduli.  When the dimension of the internal torus
is bigger than 3, then tori even of different shapes should be
identified.  It would be of interest to study the identifications in
the internal spaces in the lower-dimensional examples.

\section{$T^2$ reductions of $D=6$ gravity and supergravity}

     In sections 2 and 3, we showed that the full CJ groups for all
$D$-dimensional maximal supergravities can be described in terms of
the closure of the internal general coordinate transformations of the
toroidal compactifications of $D=11$ supergravity and of type IIB
supergravity, with type IIA/IIB T-duality linking the two pictures.
The T-duality and the general coordinate transformations $GL(10-D,\R)$
of the type IIB theory are both perturbative transformations of string
theory.  The non-perturbative aspect arises as the internal general
coordinate transformations of the toroidal reduction of 11-dimensional
supergravity.  The fact that all the non-perturbative transformations
within the CJ symmetries arise from internal general coordinate
transformations in M-theory may shed some light on the
non-perturbative aspects of string theory.

     As we mentioned in the introduction, the CJ symmetry
transformations can have the effect of altering the asymptotic
structure and also the curvature structure of a $p$-brane soliton.
Obviously, a general coordinate transformation cannot do this.  What
we have shown is that it is really the type IIA/IIB T-duality
transformation that is responsible for this change of the geometrical
structure of the $p$-brane, and hence the nature of this change may be
understandable within the perturbative framework.

     It is well known that the $R\leftrightarrow 1/R$ transformation
of type IIA/IIB T-duality is a stringy phenomenon, which becomes a
symmetry in the heterotic string, {\it i.e.}\ when the R-R fields are
truncated out.  At the level of supergravity this symmetry amounts to
an orthogonal transformation of the dilatons, accompanied by an
interchange of the Kaluza-Klein vector and the winding-mode vector
({\it i.e.}\ the vector coming from the dimensional reduction of the
antisymmetric tensor that couples to the string).  In fact this type
of symmetry, at the field theory level, can also arise in a pure
gravity theory.

    To illustrate this, let us consider pure gravity in $D=6$, and
then perform a dimensional reduction to $D=4$ on a 2-torus.  It is
straightforward to see that the four-dimensional Lagrangian has a
manifest $GL(2,\R)$ global symmetry, which is the residual symmetry of
the general coordinate transformations of the internal 2-torus.  The
scalar coset contains two dilatons $\phi_1$ and $\phi_2$, and one
axion $\chi=\cA^1_{\02}$.  The two Kaluza-Klein vectors, $\cA^1_\1$
and $\cA^2_\1$, form a doublet.  We may now show that at the level of
equations of motion, there is an additional $R\leftrightarrow 1/R$
discrete symmetry, where $R$ is the radius of the circle compactifying
from $D=5$ to $D=4$.  To this see, we first dimensionally reduce $D=6$
gravity to $D=5$, where we will obtain a dilaton $\phi_1$ and a
Kaluza-Klein vector $\cA^1_\1$, together the metric.  We now dualise
the Kaluza-Klein vector to a 2-form potential $A_\2$.  The Lagrangian
for this dualised system is given by
\be
{\cal L}_5 = R {*\oneone}-\ft12 {*d\phi_1} \wedge d\phi_1 
-\ft12 e^{\ft4{\sqrt6}\phi_1}\, {*F_\3}\wedge F_\3\ ,\label{d5lag}
\ee
where $F_\3=dA_\2$.  This theory admits an electric string soliton
solution.  The form of the Lagrangian is analogous to that for the
bosonic sector of $N=1$ supergravity in $D=10$.  Thus we expect that
upon dimensional reduction to $D=4$, there should be a discrete
symmetry that interchanges the new Kaluza-Klein vector $\cA_\1^2$ with
the winding-mode vector $A_\1$ that comes from the dimensional
reduction of $A_\2$.  Indeed, performing the reduction we find
\bea
{\cal L}_4 &=& R{*\oneone} -\ft12 {*d\vec\phi}\wedge d\vec\phi
-\ft12 e^{\ft4{\sqrt{6}}\phi_1 -\ft2{\sqrt3}\phi_2}\, {*dF_\3} \wedge
F_\3\nn\\
&& -\ft12 e^{\ft4{\sqrt6}\phi_1 + \ft1{\sqrt3}\phi_2}\, {*dF_\2}\wedge
dF_\2 -\ft12 e^{-\sqrt3\phi_2}\, {*d\cF^2_\2}\wedge
d\cF^2_\2\ ,\label{d4lag1}
\eea
where $F_\3=dA_\2 -\cA_\1^2\wedge dA_\1$, $F_\2=dA_\1$ and $\cF_\2^2 =
d\cA_\1^2$.  It is straightforward to see that (\ref{d4lag1}) is
invariant under the following discrete transformation 
\bea 
&&\pmatrix{\phi_1 \cr \phi_2} \longrightarrow \Lambda\,
\pmatrix{\phi_1\cr\phi_2} = \pmatrix{\ft13 & -\ft{2\sqrt2}3\cr
-\ft{2\sqrt2}3 &-\ft13}\pmatrix{\phi_1 \cr \phi_2}
\label{dilatontrans}\\
&&\nn\\ 
&& A_\1 \rightarrow \cA_\1^2\qquad \cA_\1^2
\rightarrow A_\1\qquad A_\2 \rightarrow A_\2 + A_\1\wedge \cA_\1^2\ .
\eea 
Note that we have $\Lambda = \Lambda^{-1}$.  If we measure the radius
of the compactifying circle from $D=5$ to $D=4$ in the
five-dimensional ``string'' metric,\footnote{If the dilaton coupling
of the 3-form field strength that couples to the fundamental string is
of the form $e^{\vec c\cdot\vec \phi}\, F_\3^2$ in the Einstein-frame
metric $ds_{\sst E}^2$ in $D$-dimensions, then the $D$-dimensional
string-frame metric is $ds_{\sst\rm str}^2 = e^{-\fft12\vec
c\cdot\vec\phi}\, ds_{\sst E}^2$.  The $D$-dimensional string coupling
constant is given by $g_{\sst D}=e^{-{\sst (D-2)/8} \, \vec c\cdot\vec
\phi}$.} {\it i.e.}\ the metric in which the mass of the 5-dimensional
string soliton is independent of the $\phi_1$ dilaton modulus, then it
is given by $R=e^{-\phi_1/\sqrt6 -\phi_2/\sqrt3}$.  This transforms as
$R\rightarrow 1/R$ under the dilaton transformation
(\ref{dilatontrans}).  Note that in the Lagrangian (\ref{d4lag1}), the
$SL(2,\R)$ factor in the $GL(2,\R)$ coming from the internal general
coordinate transformations is broken, and instead, we gain an
$R\rightarrow 1/R$ discrete symmetry.  If instead we had not dualised
the Kaluza-Klein vector in $D=5$, the resulting 4-dimensional
Lagrangian would have been invariant under $GL(2,\R)$, but then it
would not have been invariant under the discrete symmetry.  At the
level of the equations of motion, where the choice of whether or not
to dualise $\cA_\1^1$ is immaterial, the theory is invariant under
both the $GL(2,\R)$ and the discrete symmetries.

     The example we have just been considering illustrates that an
$R\leftrightarrow 1/R$ symmetry can arise in the context of a pure
gravity theory.  From the six-dimensional point of view, it
interchanges one Kaluza-Klein vector with the dual of the other
Kaluza-Klein vector,\footnote{For this reason, this $R\leftrightarrow
1/R$ symmetry lies outside the general-coordinate $GL(2,\R)$
symmetry.}  by contrast with the situation in string theory where
Kaluza-Klein and winding modes are interchanged.  Of course this
theory is just a toy model, and it is not supersymmetric.  We may,
however, embed it in a supersymmetric theory, in which case the
$R\leftrightarrow 1/R$ symmetry will be embedded as part of a larger
symmetry of the supersymmetric theory.  The minimal supergravity in
$D=6$ is $N=1$, which contains the metric and a self-dual 3-form field
strength $G_\3$.  If this is directly reduced on a 2-torus to $D=4$,
the internal $GL(2,\R)$ general coordinate transformations are
symmetries only at the level of the equations of motion, owing to the
self-duality condition on $G_\3$.  (This is analogous to the situation
for the reduction of type IIB supergravity, which is discussed in
appendix \ref{app}.)  Now, we shall follow the same strategy that we
did previously for pure gravity, and first dualise the Kaluza-Klein
vector $\cA_\1^1$ in $D=5$.  This gives the Lagrangian
\be
{\cal L}_5 = R{*\oneone} -\ft12 {*d\phi_1}\wedge d\phi_1 -
\ft12 e^{\ft4{\sqrt6}\phi_1}\, {*F_\3}\wedge F_\3 - \ft12
e^{\ft2{\sqrt6}\phi_1}\, {*G_\2}\wedge G_\2\ ,
\ee
where $G_\2=dB_\1$ comes from the dimensional reduction of $G_\3$, and
$F_\3=dA_\2 -\ft12 B_\1\wedge dB_\1$ is the field strength coming from
the dualisation of $\cF_\2^1$.  After a further reduction to $D=4$, 
followed by a dualisation of $A_\2$ to give an axion $\chi$, we obtain
the Lagrangian
\bea
{\cal L}_4 &=& R{*\oneone} -\ft12 {*d\vec\phi}\wedge d\vec\phi
-\ft12 e^{2\phi}\, {*d\chi}\wedge d\chi - \ft12 e^{\sqrt 2\varphi}\, 
{*d\sigma}\wedge d\sigma \nn\\
&&-\ft12e^{\sqrt 2\varphi-\phi}\, {*F_\2^+}\wedge F_\2^+ 
  -\ft12e^{-\phi}\, {*F_\2^0}\wedge F_\2^0
  -\ft12e^{-\sqrt 2\varphi-\phi}\, {*F_\2^-}\wedge F_\2^-\nn\\
&&-\chi\, (F_\2^+\wedge F_\2^- +\ft12 F_\2^0\wedge F_\2^0)\ ,\label{d4sdlag}
\eea
where
\be
F_\2^+ =dA_\1^+ -\sigma\, dA_\1^0 -\ft12 \sigma^2\, dA_\1^-\ ,\qquad
F_\2^0 = dA_\1^0 +\sigma\, dA_\1^-\ ,\qquad
F_\2^- = dA_\1^- \ .
\ee
The vector potential $A_\1^+$ arises from the reduction of $A_\2$ in
$D=5$; it is in fact dual to the Kaluza-Klein vector $\cA_\1^1$.  The
potential $A_\1^0$ comes from the reduction of $B_\1$ in $D=5$, and
$A_\1^-$ is the new Kaluza-Klein vector $\cA_\1^2$.  Finally, $\sigma$
is the axion coming from the reduction of $B_\1$.  The dilatons $\phi$
and $\varphi$ are related to the standard $\phi_1$ and $\phi_2$ by
\be
\phi=-\sqrt{\ft23}\, \phi_1 + \ft1{\sqrt3}\, \phi_2\ ,\qquad
 \varphi= \ft1{\sqrt3}\, \phi_1 + \sqrt{\ft23}\, \phi_2\ .
\ee
    
      The scalar sector in (\ref{d4sdlag}) contains two independent
$SL(2,\R)/O(2)$ cosets, one parameterised by $(\phi,\chi)$, and the
other by $(\varphi,\sigma)$.  We shall denote the corresponding global
symmetries by $SL(2,\R)_{\sst G}$ and $SL(2,\R)_{\sst T}$.  The
$SL(2,\R)_{\sst G}$ is a subgroup of the $GL(2,\R)$ internal
general-coordinate transformations, and hence it is a symmetry only at
the level of the equations of motion (owing to the self-duality
condition), where $F_\2^0$ and its dual form a doublet.  It is
therefore an electric/magnetic duality symmetry of the 2-form field
strength that comes from the self-dual 3-form in $D=6$.  The
$SL(2,\R)_{\sst T}$ symmetry, on the other hand, leaves the Lagrangian
(\ref{d4sdlag}) itself invariant, with the three potentials $(A_\1^+,
A_\1^0, A_\1^-)$ transforming as a triplet.  This can be made manifest
by defining the matrix
\be
{\cal M} = \pmatrix{e^{-\ft1{\sqrt2}\varphi} + \ft12 \sigma^2\, 
        e^{\ft1{\sqrt2}\varphi} & \ft1{\sqrt2}\, \sigma\,
e^{\ft1{\sqrt2}\varphi}\cr
\ft1{\sqrt2}\, \sigma\, e^{\ft1{\sqrt2}\varphi} &
 e^{\ft1{\sqrt2}\varphi} }\ ,
\ee
and the quantity
\be
A=\pmatrix{\ft1{\sqrt2}\, A_\1^0 & - A_\1^+ \cr
            - A_\1^- & -\ft1{\sqrt2}\, A_\1^0 }\ ,
\ee
in terms of which (\ref{d4sdlag}) can be written as
\bea
{\cal L}_4 &=& R{*\oneone} -\ft12{*d\phi}\wedge d\phi -\ft12
e^{2\phi}\, {*d\chi}\wedge d\chi
 +\ft12\tr({*d{\cal M}^{-1}}\wedge d{\cal M}) \nn\\
&&- \ft12 e^{-\phi}\, \tr({*H}\wedge {\cal M}\, H^T\, {\cal M}^{-1}) -
\ft12 \chi\, \tr(H\wedge H)\ ,\label{d4sdlag2}
\eea
where $H=dA$.  
     
     The Lagrangian (\ref{d4sdlag2}) is manifestly invariant under
$SL(2,\R)_{\sst T}$, whose action on the scalars and gauge potentials
is given by
\be
{\cal M} \longrightarrow \Lambda\, {\cal M}\, \Lambda^T\ ,
\qquad A\longrightarrow \Lambda\, A\, \Lambda^{-1}\ .
\ee
It is also evident from the form of the Lagrangian that its equations
of motion are invariant under $SL(2,\R)_{\sst G}$, where, as explained
earlier, the field strength $F_\2^0$ and its dual form a doublet.  In
fact under $SL(2,\R)_{\sst G}$ the other 2-form field strengths form
two further doublets, namely $(F_\2^+, {*F_\2^-})$ and
$(F_\2^-, {*F_\2^+})$.  The $R\leftrightarrow 1/R$ symmetry that we
discussed previously in the pure gravity theory is now a part of
$SL(2,\R)_{\sst T}$, corresponding to 
\be
\Lambda=\pmatrix{0 & 1\cr
                 -1 & 0}\ .
\ee
In terms of the complex field $\tau \equiv \sigma/\sqrt2 + \im \,
e^{-\varphi/\sqrt2}$, this corresponds to $\tau \leftrightarrow
-\tau^{-1}$.

   It was shown in \cite{dlr} that the 2-torus reduction of the
six-dimensional $N=2$ supergravity to $D=4$ gives a theory with
$SL(2,\R)_{\sst S}\times O(2,2)$ global symmetry.  The $O(2,2)\sim
SL(2,\R)_1\times SL(2,\R)_2$ factor is the expected T-duality symmetry
for a string theory reduced on a 2-torus; this is a symmetry of the
four-dimensional Lagrangian.  The factor $SL(2,\R)_1$ is generated by
the Kaluza-Klein axion, while $SL(2,\R)_2$ is generated by the
winding-mode axion.  The remaining $SL(2,\R)_{\sst S}$ factor is an
electric/magnetic S-duality \cite{sen1,sen2}, which is a symmetry
only at the level of the equations of motion.  In this section, we
have seen that the dimensional reduction of $N=1$ supergravity in
$D=6$ gives a theory in $D=4$ with the global symmetry $O(2,2)
\sim SL(2,\R)_{\sst G}\times SL(2,\R)_{\sst T}$. This, however, is not
the same as the $O(2,2)$ of T-duality in \cite{dlr}.  In fact, our
$SL(2,\R)_{\sst G}$ is the same as $SL(2,\R)_1$, while our
$SL(2,\R)_{\sst T}$ is the diagonal subgroup of $SL(2,\R)_{\sst S}$
and $SL(2,\R)_2$.

\section{Conclusions}

    In this paper, we have shown explicitly how the full Cremmer-Julia
symmetries of the maximal supergravities in all dimensions $3\le D\le
11$ can be interpreted in terms of general coordinate transformations
on the internal compact toroidal dimensions.  Specifically, we showed
how the combination of such general coordinate transformations in the
$D=11$ reduction picture and in the type IIB reduction picture,
intertwined by the T-duality that relates the type IIA and type IIB
theories, conspire together to generate the entire $E_{\sst{(11-D)}}$
CJ groups.  (In fact in $D=10$ type IIA, and in $D=9$, the CJ groups
{\it are} simply the internal general coordinate groups $GL(1,\R)$ and
$GL(2,\R)$ from the toroidal compactification of $D=11$ supergravity,
and it is not necessary to invoke the intertwining with the type IIB
theory in these cases.)

    The scheme that we have described allows us to interpret the
global symmetry algebras of all the maximal supergravities, with the
exception of the $D=10$ type IIB theory itself, in terms of internal
general coordinate transformations.  (Of course the global symmetry
$GL(2,\R)$ of the $S^1$ reduction of the type IIB theory {\it can} be
understood as internal general-coordinate symmetries of the $T^2$
reduction of $D=11$ supergravity.)

    Up to this point, our discussion was a classical one at the level
of the supergravity field theories. The symmetry is conjectured
\cite{ht} to survive as a discrete U-duality, in the context of string
or M-theory.  The discrete U-duality group $E_n(\Z)$ can then be
viewed as the closure of $SL(11-D,\Z)_{\rm M}$ and $SL(10-D, \Z)_{\rm
IIB}$, which are discretised subgroups of the internal general
coordinate transformations of toroidally compactified M-theory and
type IIB theory respectively.  These two groups are interwined by the
perturbative T-duality transformation of the type IIA/IIB strings.
This decomposition gives a better understanding of the origins of the
U-duality groups.  For example, it provides a natural understanding
for why sets of moduli that are related by U-duality transformations
should be identified, since any physical compactifying torus $T^n$ is
invariant under $SL(n,\Z)$ transformations.

     It should be emphasised, however, that in a given picture ({\it
i.e.}\ either M-theory or the type IIB string), the U-duality group
also involves (in $D\le8$) transformations that are not simply
internal general coordinate transformations on the torus.  The scaling
transformation of the torus, which classically is part of the internal
general coordinate transformations, is no longer a symmetry at the
quantum level in those cases ($D=10$ and $D=9$) where it remains as a
direct-product $\R$ factor in the CJ group.  (This $\R$ factor in
$GL(11-D,\R)$ cannot survive the quantisation in $D\ge 9$ since there
is no such group as $GL(n,\Z)$.)  However, in $D\le8$ it is absorbed
into the full CJ group, and transformations that change the size of
the torus do survive in the quantum discretisation to the U-duality
symmetry group.  A consequence of this is that whereas the
compactification of M-theory on $T^n$ gives different physics under
inversion of the volume of the torus when $n\le2$, it gives equivalent
physics when $n\ge 3$.

      The perturbative T-duality that relates the type IIA and type
IIB strings on circles of reciprocal radii, in the infinite-radius
limit, requires further comment.  Let us consider compactifying the
type IIA and type IIB strings on circles with radii $R_{\IIA}$
and $R_{\IIB}=1/R_{\IIA}$ respectively.  (These are measured in the
correponding ten-dimensional string metrics.)  The two theories are
equivalent, by virtue of T-duality, with the following relation
between the string coupling constants
\be
\fft{g_{10}^{\IIA}}{\sqrt{R_{\IIA}}} = g_9^{\IIA} =
g_9^{\IIB} = \fft{g_{10}^{\IIB}}{\sqrt{R_{\IIB}}}\ .
\ee
In other words, we have
\be
g_{10}^{\IIA} =\fft{g_{10}^{\IIB}}{R_{\IIB}}\ ,\quad {\rm or\,\,\,
equivalently},\quad
g_{10}^{\IIB} =\fft{g_{10}^{\IIA}}{R_{\IIA}}\ .\label{abrel}
\ee
Thus for any finite but non-vanishing value of radius $R_{\IIA}$ or
$R_{\IIB}$, both $g_{10}^{\IIA}$ and $g_{10}^{\IIB}$ can be small and
non-vanishing, and so both the type IIA and the type IIB string are
within their weak-coupling perturbative regimes.  However, in the
limit where one of the radii approaches infinity ½(or equivalently,
where the other radius becomes infinitesimal), the situation becomes
more subtle.  It follows from (\ref{abrel}) that when the
compactifying circle of one string theory shrinks until it is
comparable in size to its coupling constant, the dual string with
reciprocal compactification radius becomes strongly coupled.  For
example, if $R_{\IIA}$ goes to zero, then the type IIA string coupling
$g_{10}^{\IIA}$ must also be taken to zero in order for the theory to
map to the type IIB string at large radius at finite string coupling.
In other words, it is the free $D=9$ type IIA string on a zero-radius
circle that is T-dual to the $D=10$ type IIB string.  This free $D=9$
type IIA string can also be obtained from M-theory on a 2-torus with
vanishing volume \cite{schwarz}.  If instead we keep the string
coupling constant of type IIA (or IIB) to be fixed, and send the
corresponding radius to zero, it will map to the type IIB (or IIA)
string at infinite string coupling.  In particular, it implies that
the type IIB theory at fixed coupling on a zero-radius circle is
mapped to $D=11$ M-theory (with an infinite radius for the eleventh,
as well as for the tenth, dimension), rather than to the
weakly-coupled perturbative type IIA string in $D=10$.  This relation
between the type IIA string (or M-theory) and the type IIB string
obviously goes beyond the weak-coupling regime.

     In this paper, we also discussed the toroidal reduction of
six-dimensional pure gravity and $N=1$ supergraviity to $D=4$.  In the
former case, we found that there is an $R\leftrightarrow 1/R$ discrete
symmetry at the level of the equations of motion, in addition to the
internal general coordinate $GL(2,\R)$ symmetry.  In the case of $N=1$
supergravity, we saw that the equations of motion have an
$SL(2,\R)\times SL(2,\R)$ symmetry.

\section*{Appendix}

\appendix

\section{Type IIB reduction, and $GL(10-D,\R)$}\label{app}

    Our starting point is the Lagrangian introduced \cite{bho}, whose
equations of motion, when supplemented by the externally
(consistently) imposed self-duality constraint $H_\5={*H_\5}$, give
the equations of motion for the bosonic fields of type IIB
supergravity:
\bea
{\cal L} &=& R\, {*\oneone} -\ft12 {*d\phi}\wedge d\phi -\ft12 
e^{2\phi}\, {*d\chi}\wedge d\chi - \ft14 {*H_\5}\wedge H_\5 \nn\\
&& -\ft12 e^{-\phi}\, {*F_\3^{\ns}}\wedge F_\3^{\ns} 
-\ft12 e^{\phi}\, {*F_\3^{\rr}}\wedge F_\3^{\rr}
+\ft12 B_\4\wedge dA_\2^\ns \wedge dA_\2^\rr\ ,\label{2blag}
\eea
where 
\be
F_\3^{\ns}=dA_\2^{\ns}\ , \qquad F_\3^{\rr} =
dA_\2^{\rr} -\chi\, dA_\2^{\ns}\ ,
\qquad H_\5 =dB_\4 + \ft1{2} \ep_{pq}\, A_\2^p\wedge
dA_\2^q\ .\label{2bcs}
\ee
In the expression for $H_\5$, we are using the notation that $A_\2^\ns
=A_\2^1$, and $A_\2^\rr= A_\2^2$ here, and $\ep_{12}=1$.  Note that
$H_\5$ is normalised so that its kinetic term is one half of the
``canonical'' one.  This is because eventually we shall be
substituting the solution of the self-duality condition back into the
equations, and this will lead to a compensating doubling.

    We now reduce the Lagrangian to $D$ dimensions, using procedures
and conventions that are analogous to those used for the reduction of
$D=11$ supergravity in \cite{lpsol,cjlp1}.  Thus we reduce the metric
to $D$ dimensions using the ansatz
\be
ds_{10}^2 = e^{\vec s\cdot\vec\varphi} \, ds_{\sst D}^2
         +\sum_\a e^{2\vec\gamma_\a\cdot\vec\varphi}\, (h^\a)^2\ ,
\label{metred}
\ee
where $h^\a= dz^\a + {\cal A}_\1^\a + {\cal A}^\a_{\0 \b}\, dz^\b$.
Here, we are using indices $\a,\b,\ldots = 2,3,\ldots, 11-D$ to label
the internal compacified dimensions.  Note that we begin labelling
these with $\a=2$, corresponding to the reduction step on the
coordinate $z_2$ from $D=10$ to $D=9$.  The coordinate $z_1$ will be
reserved for the reduction of $D=11$ supergravity to $D=10$.  The
vectors $\vec s$ and $\vec\gamma_\a$ have $10-D$ components, and we
have $\vec\gamma_\a = \ft12(\vec s -\vec f_\a)$ where
\bea
\vec s &=& (s_2,s_3,\ldots, s_{11-\sst D})\ ,\nn\\
\vec f_\a &=& \Big(\underbrace{0,0,\ldots, 0}_{\a-2}, (10-\a)\, s_\a, 
s_{\a+1}, s_{\a+2}, \ldots, s_{11-\sst D} \Big)\ ,
\eea
and $s_\a= \sqrt{2/((10-\a)(9-\a))}$.  With these definitions, it
follows that a pure Einstein-Hilbert action in $D=10$ reduces to an
Einstein-Hilbert action in $D$ dimensions together with
canonically-normalised dilatonic scalars and vectors.  Note that $\vec
s$ and $\vec f_\a$ satisfy the dot-product relations
\be
\vec s\cdot\vec s= \ft{10-D}{4(D-2)}\ ,\qquad
\vec s\cdot\vec f_\a = \ft2{D-2}\ ,\qquad
\vec f_\a\cdot\vec f_\b = 2\delta_{\a\b} + \ft2{D-2}\ .
\ee

     The various potentials in (\ref{2blag}) are reduced according to
the ans\"atze
\bea
B_\4 &\rightarrow& B_\4 + B_{\3 \a}\, dz^\a +\ft12 B_{\2 \a\b}\, dz^\a\,
dz^\b + \ft16 B_{\1\a\b\g}\, dz^\a\, dz^\b\, dz^\g \nn\\
&& + \ft1{24}
B_{\0\a\b\g\d}\, dz^\a\, dz^\b\, dz^\g\, dz^\d \ ,\nn\\
A_\2 &\rightarrow& A_\2 + A_{\1 \a}\, dz^\a + \ft12 A_{\0 \a\b}\,
dz^\a\, dz^\b\ ,
\eea
where the $A$ potentials can carry either $\ns$ or $\rr$
superscripts.  Note that we are in general suppressing the $\wedge$
symbols between differential forms.  The field strengths are defined
according to the following conventions:
\bea
H_\5 &\rightarrow& H_\5 + H_{\4 \a}\, h^\a +\ft12
H_{\3 \a\b}\, h^\a\, h^\b + 
\ft16 H_{\2\a\b\g}\, h^\a\, h^\b\, h^\g \nn\\
&& + \ft1{24}
H_{\1\a\b\g\d}\, h^\a\, h^\b\, h^\g\, h^\d \ ,\nn\\
F_\3 &\rightarrow& F_\3 + F_{\2 \a}\, h^\a + \ft12 F_{\1 \a\b}\,
h^\a\, h^\b\ ,\label{h5red}
\eea
where the fields $F$ can carry $\ns$ or $\rr$ superscripts.  From
these, one can read off the expressions for the field strengths in
terms of the potentials.  For the fields coming from $H_\5$, we find
\bea
H_\5 &=& G_\4 - G_{\3 \a}\, \hA_\1^\a +\ft12 G_{\2 \a\b}\,
\hA_\1^\a \, \hA_\1^\b -\ft16 G_{\1\a\b\g}\, \hA_\1^\a\, \hA_\1^\b\,
\hA_\1^\g\nn\\
&& + \ft1{24} G_{\0\a\b\g\delta}\, \hA_\1^\a\, \hA_\1^\b\,
\hA_\1^\g\,\hA_\1^\delta\ ,\nn\\
H_{\4\a} &=&\g^\b{}_\a\Big(G_{\3 \b} + G_{\2 \b\g}\,
\hA_\1^\g +\ft12 G_{\1\b\g\delta}\, \hA_\1^\g\, \hA_\1^\delta
+ \ft1{6} G_{\0\b\g\delta\sigma}\, \hA_\1^\g\, \hA_\1^\delta\,
\hA_\1^\sigma\Big)\ ,\nn\\
H_{\3\a\b}&=& \g^\g{}_{\a}\,\g^{\delta}{}_\b\Big( G_{\2\g\delta} -
G_{\1\g\delta\sigma}\, \hA_\1^{\sigma} + \ft12
G_{\0\g\delta\sigma\rho}\, \hA_\1^{\sigma}\, \hA_\1^{\rho}\Big)\ ,\nn\\
H_{\2\a\b\g} &=& \g^{\delta}{}_\a\, \g^{\sigma}{}_\b\, \g^{\rho}{}_\g\,
(G_{\1\delta\sigma\rho} + G_{\0\delta\sigma\rho\lambda}\,
\hA^{\lambda}_\1) \ ,\nn\\
H_{\1\a\b\g\delta} &=& \g^\sigma{}_\a\, \g^\rho{}_\b\, \g^\lambda{}_\g\,
\g^\tau{}_\delta\, G_{\0\sigma\rho\lambda\tau}\ ,\label{5formred}
\eea
where 
\bea
G_\5 &=& dB_\4 +\ft12 \ep_{pq}\, A_\2^p\, dA_\2^q\ ,\nn\\
G_{\4 \a} &=& dB_{\3\a} +\ft12 \ep_{pq}\, (A_\2^p\, dA_{\1\a}^q -
A_{\1\a}\, dA_\2^q) \ ,\nn\\
G_{\3\a\b} &=& dB_{\2\a\b} +\ft12 \ep_{pq}\, (A_\2^p\, dA_{\0\a\b}^q 
+ A_{\0\a\b}\, dA_\2^q + 2 A_{\1 [\a}^p \, dA_{\1\b]}^q)\ ,\label{gdef}\\
G_{\2\a\b\g} &=& dB_{\1\a\b\g} -\ft32\ep_{pq}\, (A_{\1[\a}^p 
\, dA_{\0\b\g]}^q - A_{\0[\a\b}^p\, dA_{\1 \g]}^q)\ ,\nn\\
G_{\1\a\b\g\d} &=& dB_{\0\a\b\g\d} + 3 \ep_{pq}\, A_{\0 [\a\b}^p\, 
dA_{\0 \g\d]}^q\ .\nn
\eea
The fields coming from the NS-NS and R-R 3-forms are given by
\bea
F^\ns_\3 &=& dA^\ns_\2 - dA^\ns_{\1\a}\, \hA^{\a}_\1 + 
        \ft12 dA^\ns_{\0\a\b}\, \hA_\1^\a \, \hA_\1^\b\ ,\nn\\
F^\ns_{\2\a} &=& \gamma^\b{}_\a(dA^\ns_{\1\b} + dA^\ns_{\0\b\g}\,
\hA_\1^\g) \ ,\nn\\
F^\ns_{\1\a\b} &=& \gamma^\g{}_\a\, \g^{\delta}{}_\b\,
dA^\ns_{\0\g\delta} \ ,\nn\\
F^\rr_\3 &=& dA^\rr_\2-\chi dA^\ns_\2 - (dA^\rr_{\1\a}-\chi dA^\ns_{\1\a})
\hA^{\a}_\1 + 
        \ft12 (dA^\rr_{\0\a\b}-\chi dA^\ns_{\0\a\b}) 
               \hA_\1^\a \, \hA_\1^\b\ ,\nn\\
F^\rr_{\2\a} &=& \gamma^\b{}_\a\, \Big(dA^\rr_{\1\b} -\chi dA^\ns_{\1\b}
   + (dA^\rr_{\0\b\g}-\chi dA^\ns_{\0\b\g}) \hA_\1^\g \Big) \ ,\nn\\
F^\rr_{\1\a\b} &=& \gamma^\g{}_\a\, \g^{\delta}{}_\b
(dA^\rr_{\0\g\delta} -\chi dA^\ns_{\0\g\delta}) \ .\label{3formsred}
\eea
In these expressions, the quantity $\gamma^\a{}_\b$ is the inverse of
$\tilde\gamma^\a{}_\b \equiv \delta^\a_\b +{\cal A}^\a_{\0\b}$, 
and $\hA_\1^\a\equiv \g^\a{}_\b\, {\cal A}_\1^\b$ (see \cite{lpsol,cjlp1}).
Note that $dz^\a = \gamma^\a{}_\b\, h^\b - \hA_\1^\a$.

    It is now straightforward to calculate the $D$-dimensional
Lagrangian following from (\ref{2blag}).  We find
\bea
{\cal L} &=& R\, {*\oneone} -\ft12 { *d\vec\phi}\wedge d\vec\phi
   -\ft12 e^{\vec d\cdot\vec\phi}\, {*d\chi}\wedge d\chi  
-\ft14 e^{\vec c\cdot\vec\phi}\, {*H_\5}\wedge H_\5 -
\ft14 \sum_\a e^{\vec c_\a\cdot\vec\phi}\, {*H_{\4 \a}}\wedge H_{\4\a}\nn\\
&&-\ft14\sum_{\a <\b} e^{\vec c_{\a\b}\cdot\vec\phi}\, 
                       {*H_{\3 \a\b}}\wedge H_{\3\a\b}
-\ft14\sum_{\a <\b<\g } e^{\vec c_{\a\b\g}\cdot\vec\phi}\,
                       {*H_{\2 \a\b\g}}\wedge H_{\2\a\b\g}\nn\\
&&
-\ft14\sum_{\a <\b<\g <\d} e^{\vec c_{\a\b\g\d}\cdot\vec\phi}\,
                       {*H_{\1 \a\b\g\d}}\wedge H_{\1\a\b\g\d}\label{dlag}\\
&&-\ft12 e^{\vec a^\ns\cdot\vec\phi}\, {*F_\3^\ns}\wedge F_\3^\ns
-\ft12\sum_\a e^{\vec a^\ns_\a\cdot\vec\phi}\, 
                 {*F_{\2\a}^\ns}\wedge F_{\2\a}^\ns
-\ft12\sum_{\a<\b} e^{\vec a^\ns_{\a\b}\cdot\vec\phi}\,
                 {*F_{\1\a\b}^\ns}\wedge F_{\1\a\b}^\ns\nn\\
&&-\ft12 e^{\vec a^\rr\cdot\vec\phi}\, {*F_\3^\rr}\wedge F_\3^\rr
-\ft12\sum_\a e^{\vec a^\rr_\a\cdot\vec\phi}\,
                 {*F_{\2\a}^\rr}\wedge F_{\2\a}^\rr
-\ft12\sum_{\a<\b} e^{\vec a^\rr_{\a\b}\cdot\vec\phi}\,
                 {*F_{\1\a\b}^\rr}\wedge F_{\1\a\b}^\rr\nn\\
&&-\ft12 \sum_\a e^{\vec b_\a\cdot \vec\phi}\, {*{\cal F}_\2^\a}
\wedge {\cal F}_\2^\a -\ft12 \sum_{\a <\b} e^{\vec b_{\a\b}\cdot\vec
\phi}\, {*{\cal F}^\a_{\1\b}}\wedge {\cal F}^\a_{\1\b}  
   +{\cal L}_{\sst{FFB}}\ .\nn
\eea
Here, $\vec\phi\equiv(\phi,\vec\varphi)$, where $\vec\varphi$ denotes the 
set of dilatonic scalars appearing in (\ref{metred}) that come from
the dimensional reduction from 10 to $D$ dimensions, and $\phi$ is the
original dilaton of the ten-dimensional type IIB theory.  The various
dilaton vectors for the $D$-dimensional field strengths are given by
\bea
&& \vec a^\ns =(-1, -4\vec s) \ ,\qquad 
   \vec a^\ns_\a =(-1, -4\vec s+ \vec f_\a ) \ ,\qquad
   \vec a^\ns_{\a\b} =(-1, -4\vec s+ \vec f_\a + \vec f_\b) \ ,\nn\\
&& \vec a^\rr =(1, -4\vec s) \ ,\qquad
   \vec a^\rr_\a =(1, -4\vec s+ \vec f_\a ) \ ,\qquad
   \vec a^\rr_{\a\b} =(1, -4\vec s+ \vec f_\a + \vec f_\b) \ ,\nn\\
&&
    \vec b_\a = (0,-\vec f_\a)\ ,\qquad 
    \vec b_{\a\b} = (0,-\vec f_\a + \vec f_\b)\ ,\qquad
    \vec d =(2,\vec 0)\ ,\nn\\
&&  \vec c= (0,-8\vec s) \ ,\qquad
    \vec c_\a= (0,-8\vec s + \vec f_\a) \ ,\qquad
    \vec c_{\a\b}= (0,-8\vec s + \vec f_\a + \vec f_\b) \ ,\nn\\
&&  \vec c_{\a\b\g}= (0,-8\vec s + \vec f_\a + \vec f_\b+\vec f_\g) \ ,\qquad
  \vec c_{\a\b\g\d}= (0,-8\vec s + \vec f_\a + \vec f_\b+\vec f_\g
                            +\vec f_\d) \ .
\eea
Note that ${\cal L}_{\sst{FFB}}$ represents the Chern-Simons terms,
coming from the dimensional reduction of the last term in
(\ref{2blag}).  They are straightforwardly calculated using the
ans\"atze for the reduction of the potentials, and we find
\bea
D=10:&&\ft12 B_\4\wedge dA^{\ns}_\2 \wedge dA^{\rr}_\2\ ,\nonumber\\
D=9: &&\ft12\Big(\ep_{pq}\, B_\4 \wedge dA^p_\2 \wedge dA^q_{\1 2}
 + B_{\3 2} \wedge dA^{\ns}_\2 \wedge dA^{\rr}_\2\Big),\nonumber\\
D=8: && \Big(\ft14\ep_{pq}\, B_\4\wedge dA^p_\2\wedge dA^q_{\0\a\b}
+ \ft12B_\4\wedge dA^{\ns}_{\1\a}\wedge dA^{\rr}_{\1\b}\nonumber\\
&& -\ft12 \ep_{pq}\, B_{\3\a}\wedge dA^p_\2\wedge dA^q_{\1\b}
+ \ft14 B_{\2\a\b}\wedge dA^{\ns}_\2 \wedge
dA^{\rr}_\2\Big)\epsilon^{\a\b}
\ ,\nonumber\\
D=7:&&
\Big(-\ft14\ep_{pq}\,  B_\4\wedge dA^p_{\1\a}\wedge dA^q_{\0\b\g}
+ \ft12 B_{\3\a}\wedge dA^{\ns}_{\1\b}\wedge dA^{\rr}_{\1\g}\nonumber\\
&&+ \ft14\ep_{pq}\,  B_{\3\a}\wedge dA^p_\2\wedge dA^q_{\0\b\g}
+ \ft14 \ep_{pq}\, B_{\2\a\b}\wedge dA^p_\2 \wedge
dA^q_{\1\g}\nn\\
&& + \ft1{12} B_{\1\a\b\g}\wedge dA^\ns_\2\wedge dA^\rr_\2
\Big)\epsilon^{\a\b\g}\ ,\nonumber\\
D=6: && \Big(
\ft14 \ep_{pq}\, B_{\3\a}\wedge dA^p_{\1\b}\wedge dA^q_{\0\g\d}
+\ft12 B_{\2\a\b}\wedge dA^{\ns}_{\1\g}\wedge dA^{\rr}_{\1\d}\nn\\
&&+\ft18 \ep_{pq}\, B_{\2\a\b}\wedge dA^p_\2\wedge dA^q_{\0\g\d}
+\ft18 B_\4\wedge dA^\ns_{\0\a\b}\wedge dA^\rr_{\0\g\d}\nn\\
&&-\ft1{12}\ep_{pq}\,  B_{\1\a\b\g}\wedge dA^p_\2\wedge dA^q_{\1\d}
+\ft1{48} B_{\0\a\b\g\d}\wedge dA^{\ns}_\2\wedge
dA^{\rr}_\2\Big)\epsilon^{\a\b\g\d}, \nonumber\\
D=5: &&
\Big(
-\ft18 B_{\3\a}\wedge dA^{\ns}_{\0\b\g}\wedge dA^{\rr}_{\0\d\e} -
\ft18\ep_{pq}\,  B_{\2\a\b}\wedge dA^p_{\1\g}\wedge dA^q_{\0\d\e}\nonumber \\
&&-\ft1{24}\ep_{pq}\,  B_{\1\a\b\g}\wedge dA^p_\2\wedge dA^q_{\0\d\e}+
\ft1{12} B_{\1\a\b\g}\wedge dA^{\ns}_{\1\d}\wedge dA^{\rr}_{\1\e}\nonumber\\
&&+\ft1{48}\ep_{pq}\, B_{\0\a\b\g\d}\wedge dA^p_\2\wedge dA^q_{\1\e}
\Big)\epsilon^{\a\b\g\d\e}, \nonumber\\
D=4:&&\Big(
\ft1{16} B_{\2\a\b}\wedge dA^{\ns}_{\0\g\d}\wedge dA^{\rr}_{\0\e\z}
+ \ft1{96}\ep_{pq}\, B_{\0\a\b\g\d}\wedge dA^p_\2\wedge dA^q_{\0\e\z}
\nonumber\\
&&+ \ft1{48} B_{\0\a\b\g\d}\wedge dA^{\ns}_{\1\e}\wedge dA^{\rr}_{\1\z}+
+\ft1{24} \ep_{pq}\, B_{\1\a\b\g}\wedge dA^p_{\1\d}\wedge dA^q_{\0\e\z}
\Big)\epsilon^{\a\b\g\d\e\z}\ ,\nonumber\\
D=3:&& \Big(
\ft1{48} B_{\1\a\b\g}\wedge dA^{\ns}_{\0\d\e}\wedge dA^{\rr}_{\0\z\h}
- \ft1{96}\ep_{pq}\, B_{\0\a\b\g\d}\wedge dA^p_{\1\e}\wedge
dA^q_{\0\z\h}
\Big)\epsilon^{\a\b\g\d\e\z\h}, \nonumber\\
D=2: &&
\ft1{192} B_{\0\a\b\g\d}\wedge dA^{\ns}_{\0\e\z}\wedge dA^{\rr}_{\0\h\q}
\, \epsilon^{\a\b\g\d\e\z\h\q}\ .\label{ffb}
\eea
(Recall that we are sometimes using the notation $A^p$ with $p=1$ or $p=2$
to indicate $A^\ns$ or $A^\rr$ respectively.)

    The $D$-dimensional Lagrangian (\ref{dlag}) has a global
$GL(10-D,\R)$ invariance.  In particular, the various fields with
$\a,\b,\ldots$ indices each transform irreducibly under $GL(10-D,\R)$,
in the corresponding antisymmetric-tensor representations.
Furthermore, it is evident from the invariance of the Lagrangian that
the duals of the $H$ fields, defined by
\bea
 &&\tH_{\sst{(D-5)}}\equiv  e^{\vec c\cdot\vec\varphi}\, {*H_\5}\ ,\qquad
\tH_{\sst{(D-4)}}^\a \equiv e^{\vec c_\a\cdot\vec\varphi}\, {*H_{\4
\a}}\ ,\qquad
\tH_{\sst{(D-3)}}^{\a\b} \equiv e^{\vec c_{\a\b}\cdot\vec\varphi}\, {*H_{\3
\a\b}}\ ,\nn\\
&&\tH_{\sst{(D-2)}}^{\a\b\g} \equiv e^{\vec
c_{\a\b\g}\cdot\vec\varphi}\, {*H_{\2 \a\b\g}}
\ ,\qquad \tH_{\sst{(D-1)}}^{\a\b\g\d}
     \equiv e^{\vec c_{\a\b\g\d}\cdot\vec\varphi}\, {*H_{\1
 \a\b\g\d}}\ ,
\label{dualh}
\eea
transform covariantly as antisymmetric-tensor representations with upstairs
$GL(10-D,\R)$ indices.  

   We now consider the self-duality condition which must be imposed on
the ten-dimensional 5-form field strength.  It is useful to do this by
postponing the imposition of the self-duality constraint until the
lower dimension $D$ has been reached.  At this stage, the original
condition $H_\5={*H_\5}$ becomes a set of conditions relating the
various dimensionally-reduced components of the original $H_\5$.  To
carry out this procedure, it is useful to define the following forms
in the internal space:
\bea
\Sigma &=& \ft1{(10-D)!}\, \epsilon_{\b_1 b_2\cdots
\b_{\sst{(10-D)}}}\,
             h^{\b_1} \wedge h^{\b_2}\wedge \cdots \wedge
             h^{\b_{\sst{(10-D)}}}\ ,\nn\\
\Sigma_{\a_1} &=& \ft1{(9-D)!}\,
           \epsilon_{\a_1 \b_1\cdots \b_{\sst{(9-D)}}}\,
             h^{\b_1}\wedge \cdots \wedge
             h^{\b_{\sst{(9-D)}}}\ ,\label{sigdef}\\
\Sigma_{\a_1 \a_2} &=& \ft1{(8-D)!}\,
           \epsilon_{\a_1 \a_2 \b_1\cdots \b_{\sst{(8-D)}}}\,
             h^{\b_1}\wedge \cdots \wedge
             h^{\b_{\sst{(8-D)}}}\ .\nn\\
& \cdots &\nn
\eea
We then find that the dimensional reduction of the ten-dimensional
Hodge dual of $H_\5$ gives
\be
(-1)^{\sst D}\, {*H _\5} \rightarrow  
\tH_{\sst{(D-5)}}\, \Sigma -
\tH_{\sst{(D-4)}}^\a \, \Sigma_\a +
\ft12 \tH_{\sst{(D-3)}}^{\a\b}\, \Sigma_{\a\b} 
-\ft16 \tH_{\sst{(D-2)}}^{\a\b\g}\, \Sigma_{\a\b\g}
+\ft1{24} \tH_{\sst{(D-1)}}^{\a\b\g\d}\, \Sigma_{\a\b\g\d}\ ,
\ee
where the tilded fields are defined in (\ref{dualh}).
From this and  (\ref{h5red}), we see that the ten-dimensional
self-duality condition $H_\5 = {*H_\5}$ becomes, in the various lower
dimensions,
\bea
D=9: && \tH_\4 = H_{\4 2}\ ,\nn\\
D=8: && \tH_\3 = \ft12 \ep^{\a\b}\,  H_{\3\a\b}\ ,\qquad
       \tH_\4^\a = -\ep^{\a\b}\, H_{\4\b}\ ,\nn\\
D=7: && \tH_\2 = -\ft16 \ep^{\a\b\g}\, H_{\2\a\b\g}\ ,\qquad
        \tH_\3^\a =\ft12 \ep^{\a\b\g}\, H_{\3\b\g}\ ,\nn\\
D=6: && \tH_\1 =\ft1{24}\, \ep^{\a\b\g\d}\, H_{\1\a\b\g\d}\, \qquad
        \tH_\2^\a = -\ft16 \ep^{\a\b\g\d}\, H_{\2\b\g\d}\ ,\qquad
        \tH_\3^{\a\b} = \ft12 \ep^{\a\b\g\d}\, H_{\3\g\d}\ ,\nn\\
D=5: && \tH_\1^\a =-\ft1{24}\, \ep^{\a\b\g\d\sigma}\,
    H_{\1\b\g\d\sigma}\ ,\qquad \tH_\2^{\a\b} = \ft16
\ep^{\a\b\g\d\sigma}\, H_{\2 \g\d\sigma}\ ,\nn\\
D=4: && \tH_\1^{\a\b} = \ft1{24} \ep^{\a\b\g\d\sigma\rho}\, 
   H_{\1 \g\d\sigma\rho}\ ,\qquad
    \tH_\2^{\a\b\g} = -\ft16 \ep^{\a\b\g\d\sigma\rho}\, 
    H_{\2 \d\sigma\rho}\ ,\nn\\
D=3: && \tH_\1^{\a\b\g} = -\ft1{24} \ep^{\a\b\g\d\sigma\rho\lambda} \,
     H_{\1 \d\sigma\rho\lambda}\ .\label{lowersd}
\eea 
It should be emphasised that these conditions are all $GL(10-D,\R)$
covariant, and so at the level of the equations of motion, the full
$GL(10-D,\R)$ general coordinate symmetry from the reduction on the
$(10-D)$-torus is manifest.  It is possible to use these equations to
substitute back into (\ref{dlag}), and thereby eliminate the doubling
of the degrees of freedom associated with the original non-self-dual
$H_\5$ field in $D=10$.  When $D$ is odd, this can be done in a
completely $GL(10-D,\R)$-covariant way, by using the individual
equations in (\ref{lowersd}) to eliminate sets of either $H$ or $\tH$
fields.  When $D$ is even, the explicit elimination of redundant
fields requires a breaking of $GL(10-D,\R)$ to $GL(9-D,\R)$, since one
has to eliminate half of the set of fields of degree $D/2$.

\end{document}